\documentclass[oldversion]{aa}  
\usepackage[pdftex]{graphicx}
\usepackage{epstopdf}
\usepackage{txfonts}
\usepackage{natbib}
\bibpunct{(}{)}{;}{a}{}{,} % to follow the A&A style
\begin{document}
   \title{Star formation efficiency in galaxy interactions and mergers:\\ a statistical study}

%   \subtitle{I. Overviewing the $\kappa$-mechanism}

   \author{P. Di Matteo
%          \inst{1}
          \and
          F. Combes%\inst{1}
          \and
          A.-L.Melchior%\inst{1}
          \and
          B. Semelin%\inst{1}
          }

   \offprints{P. Di Matteo}

   \institute{Observatoire de Paris, LERMA, 61, Avenue de L'Observatoire, 75014 Paris, France\\
%              \email{paola.dimatteo@obspm.fr; francoise.combes@obspm.fr; \\  anne-laure.melchior@obspm.fr; benoit.semelin@obspm.fr}
             }

   \date{Received ; Accepted }

% \abstract{}{}{}{}{} 
% 5 {} token are mandatory
 
  \abstract{We investigate the enhancement of star formation efficiency in galaxy interactions
and mergers, by numerical simulations of several hundred galaxy collisions.
All morphological types along the Hubble sequence are considered in the initial
conditions of the two colliding galaxies, with varying bulge-to-disk ratios and gas
mass fractions. Different types of orbits are simulated, direct and retrograde,
according to the initial relative energy and impact parameter, and the
resulting star formation history is compared to that occuring in the two
galaxies when they are isolated. Our principal results are:
(1) retrograde encounters have a larger star formation efficiency (SFE) than direct
encounters, 
(2) the amount of gas available in the galaxy is not the main parameter
governing the SFE in the burst phase,
(3) there is an anticorrelation between the amplitude of the star forming
burst and the tidal forces exerted per unit of time, which is due to the 
large amount of gas dragged outside the galaxy by tidal tails in strong interactions;
(4) globally, the Kennicutt-Schmidt law is retrieved statistically for isolated galaxies, interacting pairs and mergers;
%for mergers, but the slope of the SFR to gas surface density relation is steeperfor galaxy flybys;
(5) the enhanced star formation is essentially occurring in nuclear starbursts,
triggered by inward gas flows driven by non-axisymmetries in the galaxy disks.
Direct encounters develop more pronounced asymmetries than retrograde ones.
Based on these statistical results, we derive general laws for the enhancement
of star formation in galaxy interactions and mergers, as a function of the
main parameters of the encounter.}

   \keywords{}

\maketitle
%
%________________________________________________________________

\section{Introduction}
Theories about formation and evolution of galaxies have undergone
drastic changes in the last decade. Impressive progress in the
observations of galaxies at high and intermediate redshift now
put constraints directly on galaxy evolution \citep{ste99,lef00,cim02,flo04}.
The star formation history
can be estimated \citep{mad96,lil96}, and  the
confrontation with numerical simulations help to discover the
actual mechanisms for mass assembly and transformation of gas
into stars \citep{kau98, som01}.
 Models are done on one hand through cosmological N-body simulations
 \citep[e.g.][]{spr03,som03,gov04}
but in general they lack  spatial resolution on galaxy scales to resolve star-formation related processes in a realistic way; %and
%star formation processes are dealt with very approximately and globally;
on the other hand, simulations are then completed by semi-analytic
computations, where galaxy evolution is treated with recipes
taking into account all important processes, including hierarchical
merging and star formation \citep{dev00,kau00}.

In both cases, it is necessary to study in more details the star
formation efficiency on galaxy scale, to calibrate the physical
recipes: star formation rate, and feedback processes have
been computed on isolated and merging galaxies,
and a large variety of results have been obtained
\citep{mih94,mih94b,mih96,spri00,tis02,mez03,kap05,cox06}.
 All studies demonstrated that galaxy collisions trigger
star formation, and that the initial disk stability is the main parameter
to influence the star formation sequence: late-type galaxies without bulge
are more prone to violent bar instability during an encounter, which
drives the internal gas towards the galaxy center to trigger a nuclear
starburst \citep{mih96}. However, the availability of
gas in the interacting galaxies is also one of the most determining
parameter, as well as the adopted gas physics \citep{cox05}.

The amount of triggered star formation due to the interaction is
a fundamental parameter required by semi-analytical simulations, that
strongly depends on models adopted for star formation rate and feedback.
We propose here to investigate this problem statistically, simulating
all galaxy types along the Hubble sequence, and exploring all physical parameters
for the interactions, mass ratios, geometrical orbital parameters, etc.. in order
to get an insight into the global phenomena, and their range of variations.
In this first paper, the numerical method and galaxy models are presented,
and star formation evolution involving giant-like galaxies, are presented
and discussed. A wider range of interactions and mergers
involving the whole mass spectrum of galaxies will be considered
in a companion paper. Galaxies are initiated at z=0, while simulations involving
galaxies at any redshift will be studied in a following work.\\

The scheme of the paper is as following: in Section \ref{ini}, the adopted galaxy models (Section \ref{galmod}) and the initial orbital parameters (Section \ref{orbital}) are described; the numerical code is described in Section \ref{Numericalmethods}; in Section \ref{results}, the main results are presented, in particular those related to the evolution of the star formation rate during the different phases of interaction, its dependence on the total gas amount in the galaxies and on the main orbital parameters of the encounter.
Finally, in Section \ref{discussion} we analyze and discuss the interactions-starbursts connection and in Section \ref{concl} the main conclusions of this work are drawn.\\

%__________________________________________________________________

\section{Initial conditions}\label{ini}
\subsection{Galaxy models: moving along the Hubble sequence}\label{galmod}
%table 1
   \begin{table}
      \caption[]{Galaxy parameters. The bulge and the halo are modelled as Plummer spheres, with characteristic masses $M_B$ and $M_H$ and characteristic radii $r_B$ and $r_H$.  $M_{*}$ and  $M_{g}$ represent the masses of the stellar and gaseous disks, whose vertical and radial scale lengths are given, respectively, by $h_{*}$ and $a_{*}$, and $h_{g}$ and $a_{g}$.}      
         \label{galpar}
	 \centering
         \begin{tabular}{lcccc}
            \hline\hline
	    & gE0 & gSa & gSb & gSd \\
            \hline
	    $M_{B}\ [2.3\times 10^9 M_{\odot}]$ & 70 & 10 & 5 & 0 \\
	    $M_{H}\ [2.3\times 10^9 M_{\odot}]$ & 30 & 50 & 75 & 75\\
	    $M_{*}\ [2.3\times 10^9 M_{\odot}]$ & 0 & 40 & 20 & 25\\
	    $M_{g}/M_{*}$ & 0 & 0.1 & 0.2 & 0.3\\
	    & & & & \\
	    $r_{B}\ [\mathrm{kpc}]$ & 4 & 2 & 1& --\\
	    $r_{H}\ [\mathrm{kpc}]$ & 7 & 10 & 12 & 15\\
	    $a_{*}\ [\mathrm{kpc}]$ & -- & 4 & 5 & 6\\
	    $h_{*}\ [\mathrm{kpc}]$ & -- & 0.5 & 0.5 & 0.5\\
	    $a_{g}\ [\mathrm{kpc}]$ & -- & 5 & 6 & 7\\
	    $h_{g}\ [\mathrm{kpc}]$ & -- & 0.2 & 0.2 & 0.2\\
            \hline
         \end{tabular}
   \end{table}

\begin{figure}
   \centering
   \includegraphics[width=2.2cm,angle=270]{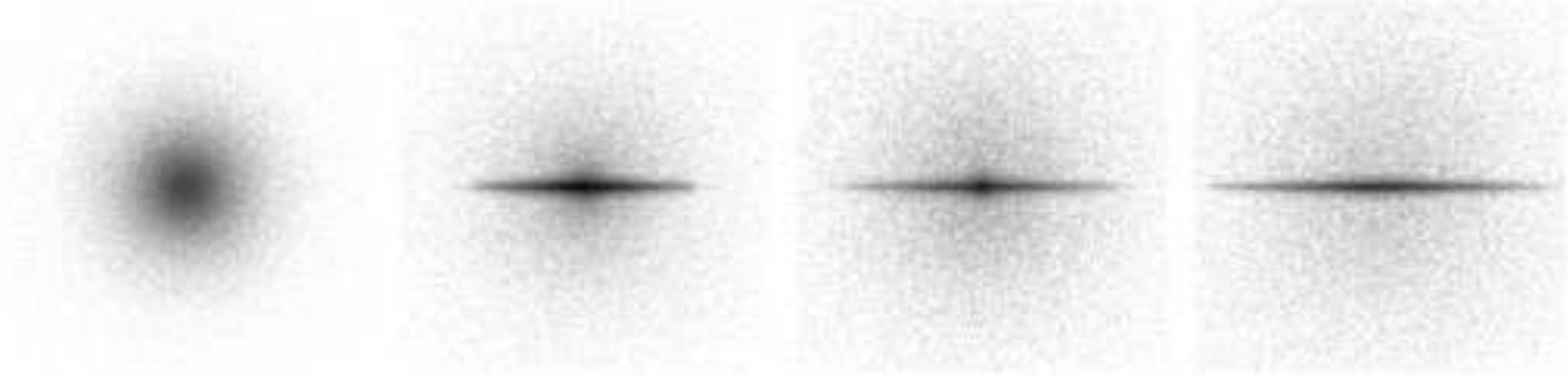}
   \caption{Hubble sequence for our galaxy models. From left to right, projection on the x-z plane of a gE0, gSa, gSb and gSd galaxy. All the different components (gas, stars and dark matter) are plotted. Dark shading represents densest regions. Each frame is 20 kpc $\times$ 20 kpc in size.}
              \label{hubble}%
    \end{figure}

\begin{figure}
   \centering
   \includegraphics[width=6.cm,angle=270]{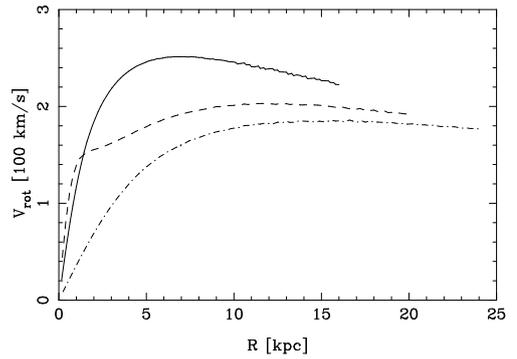}
   \caption{Initial rotation curves for the gSa (solid line), gSb (dashed line) and gSd (dot-dashed line) galaxy.}
              \label{rot}%
    \end{figure}

Aiming at exploiting a large set of interactions, involving \emph{galaxies of all morphologies from ellipticals to late-type spirals}, the galaxy models adopted consist in a spherical dark matter halo, containing or not a stellar and a gaseous disk and, optionally, a central bulge. \\
For each galaxy type\footnote{Hereafter we will adopt the following nomenclature for the different morphological types: gE0 for giant-like ellipticals, gSa for giant-like Sa spirals, gSb for giant-like Sbc spirals and gSd for giant-like Sd spirals.}, the halo and the optional bulge are modeled as a Plummer sphere \citep[][pag.42]{bt1}, with characteristic masses $M_B$ and $M_H$ and characteristic radii $r_B$ and $r_H$. Their  densities are given, respectively, by:
\begin{equation}\label{halo}
\rho_{H}(r)=\left(\frac{3M_{H}}{4\pi {r_{H}}^3}\right)\left(1+\frac{r^2}{{r_{H}}^2}\right)^{-5/2}
\end{equation}
and
\begin{equation}\label{bulge}
\rho_{B}(r)=\left(\frac{3M_{B}}{4\pi {r_{B}}^3}\right)\left(1+\frac{r^2}{{r_{B}}^2}\right)^{-5/2}.
\end{equation}
The stellar and gaseous disks follow a Miyamoto-Nagai density profile \citep[][pag.44]{bt1}: 
\begin{eqnarray}\label{stdisk}
\rho_{*}(R,z)&=&\left(\frac{{h_{*}}^2 M_{*}}{4 \pi}\right)\times\nonumber\\&&\frac{a_{*} R^2+(a_{*}+3\sqrt{z^2+{h_{*}}^2})\left(a_{*}+\sqrt{z^2+{h_{*}}^2}\right)^2}
{\left[a_{*}^2+\left(a_{*}+\sqrt{z^2+{h_{*}}^2}\right)^2\right]^{5/2}\left(z^2+{h_*}^2\right)^{3/2}}
\end{eqnarray}
\begin{eqnarray}\label{gasdisk}
\rho_{g}(R,z)&=&\left(\frac{{h_{g}}^2 M_{g}}{4 \pi}\right)\times\nonumber\\&&\frac{a_{g} R^2+(a_{g}+3\sqrt{z^2+{h_{g}}^2})\left(a_{g}+\sqrt{z^2+{h_{g}}^2}\right)^2}
{\left[a_{g}^2+\left(a_{g}+\sqrt{z^2+{h_{g}}^2}\right)^2\right]^{5/2}\left(z^2+{h_g}^2\right)^{3/2}},
\end{eqnarray}
with masses $M_{*}$ and  $M_{g}$ and  vertical and radial scale lengths given, respectively, by $h_{*}$ and $a_{*}$, and $h_{g}$ and $a_{g}$.
Moving along the Hubble sequence, from giant-like ellipticals (gE0) to giant-like Sd spiral (gSd), the mass of the central spheroid varies from $M_B=1.6\times10^{11} M_{\odot} $ for a gE0 to $M_B=0$ for a gSd, while the gas mass  $M_g$, absent in the case of a gE0, increases from $9.2\times 10^9 M_{\odot}$ in a gSa to $1.7\times10^{10} M_{\odot}$ for a gSd (see Table \ref{galpar} for a complete list of all the parameters in Eqs.\ref{halo} to \ref{gasdisk} and  Fig.\ref{hubble} for a representation of our galaxy sequence). \\The initial rotation curves for spiral galaxies are shown in Fig.\ref{rot}.  In accordance with observations \citep{rob94}, for a fixed distance from the galaxy center, the value of $V_{rot}$ is higher for early type systems than for late type ones, indicating a decrease in the enclosed mass content. Aiming at investigating interactions between giant like galaxies, the mass ratios of the interacting systems is always of order unity. In the next future, we will exploit collisions involving galaxies of the whole mass spectrum. \\

%fig1

%table 2
   \begin{table}
      \caption[]{Particle numbers for each galactic composant}
         \label{numbers}
	 \centering
         \begin{tabular}{lcccc}
            \hline\hline
	    & gE0 & gSa & gSb & gSd \\
            \hline
	    $N_{gas}$ & -- & 20000 & 40000 & 60000\\
	    $N_{star}$ & 80000 & 60000 & 40000 & 20000\\
	    $N_{DM}$ &40000 & 40000 & 40000 & 40000\\
            \hline
         \end{tabular}
   \end{table}

Intending to obtain hundreds of simulations, each galaxy is made up of 120000 particles, distributed among gas, stars and dark matter, depending on the morphological type (see Table \ref{numbers}). 

To initialize particle velocities, we adopted the method described in \citet{hern93}.

\subsection{Orbital parameters}\label{orbital}

%table 3
   \begin{table}
      \caption[]{Galaxies orbital parameters}
         \label{orbpos}
	 \centering
         \begin{tabular}{cccccc}
            \hline\hline
%	    id & $r_{ini}\ \mathrm{[kpc]}$ & ${r_{p}}^{\mathrm{a}}\ \mathrm{[kpc]}$& ${v_{p}}^{\mathrm{a}}\ \mathrm{[10^2kms^{-1}]}$ &${E}^{\mathrm{a,b}}\ \mathrm{[10^4km^2s^{-2}]}$ \\
	    id & $r_{ini}$ & ${r_{p}}^{\mathrm{a}}\ $& ${v_{p}}^{\mathrm{a}}\ $ &${E}^{\mathrm{a,b}}\ $ & spin$^{\mathrm{c}}$\\
	     & $ \mathrm{[kpc]}$ & $ \mathrm{[kpc]}$& $ \mathrm{[10^2kms^{-1}]}$ & $\mathrm{[10^4km^2s^{-2}]}$ &\\
            \hline
	    01dir &  100. & 8.0 &  7.07 &0.0 & up\\
	    01ret &  100.&  8.0& 7.07& 0.0 & down\\
	    02dir &  100.&   8.0&  7.42 & 2.5& up\\
	    02ret &  100.&  8.0& 7.42& 2.5 & down\\
	    03dir &100. &  8.0 & 7.74& 5.0& up\\
	    03ret &  100. &   8.0 & 7.74&5.0 & down\\
	    04dir &100. &   8.0 & 8.94& 15.0 & up\\
	    04ret &  100. &   8.0 & 8.94& 15.0 & down\\
	    05dir& 100. & 16.0&  5.00& 0.0 & up\\
	    05ret & 100.&16.0&  5.00& 0.0 & down\\
	    06dir & 100.  & 16.0 &  5.48&2.5 & up\\
	    06ret & 100. & 16.0 &  5.48 & 2.5 & down\\
	    07dir &100.& 16.0&  5.92& 5.0 & up\\
	    07ret & 100.& 16.0&  5.92& 5.0 & down\\
	    08dir& 100.& 16.0& 7.42& 15.0 & up\\
	    08ret&100.& 16.0& 7.42& 15.0 & down\\
	    09dir& 100.& 24.0&  4.08 &0.0 & up\\
	    09ret&100.& 24.0&  4.08 &0.0 & down\\
	    10dir& 100.& 24.0&  4.65 &2.5 & up\\
	    10ret&100.&24.0&  4.65 &2.5 & down\\
	    11dir& 100.& 24.0&  5.16 &5.0 & up\\
	    11ret&100.& 24.0&  5.16 &5.0 & down\\
	    12dir& 100.& 24.0&  6.83 &15.0 & up\\
	    12ret&100.& 24.0&  6.83 &15.0 & down\\
            \hline
         \end{tabular}

\begin{list}{}{}
\item[$^{\mathrm{a}}$] For two equal point masses of mass $m=2.3\times10^{11}M_{\odot}$.
\item[$^{\mathrm{b}}$] It is the total energy of the relative motion, i.e.\\ $E={v}^2/2-G(m_1+m_2)/r$.
\item[$^{\mathrm{c}}$] Orbital spin, if parallel (up) or antiparallel (down) to the galaxies spin.
\end{list}
   \end{table}

Aiming at exploiting a vast range of orbital parameters, for each couple of interacting galaxies we performed 24 different simulations, varying the galaxies orbital initial conditions, in order to have (for the ideal Keplerian orbit of two equal point masses of mass $m=2.3\times10^{11}M_{\odot}$) the first pericenter separation $r_{per}=$8, 16, and 24 kpc. For each of these separations, we varied the relative velocities at pericenter, in order to have one parabolic and three hyperbolic orbits of different energy. Finally, for each of the selected orbits, we changed the sign of the orbital angular momentum in order to study both direct and retrograde encounters.\\
Combining each orbital
configuration with all possible morphologies for the interacting pair of
galaxies, we obtained a total sample of 240 interactions. \\
In Table \ref{orbpos}, the initial distance $r_{ini}$ and the pericenter distance $r_{p}$ between the galaxies center-of-mass are listed, together with their relative velocity $v_{p}$ at pericenter and the orbital energy $E$,  for all the simulated encounters\footnote{The values refer to the ideal Keplerian orbit of  two equal point masses of mass $m=2.3\times10^{11}M_{\odot}$.}.\\
The results described in this paper refer only to planar encounters (galactic disks lying in the orbital plane).

\section{Numerical method}\label{Numericalmethods}

To model galaxy evolution, we employed a Tree-SPH code, in which gravitational forces are calculated using a hierarchical tree method \citep{bh86} and gas evolution is followed by means of smoothed particle hydrodynamics \citep{lucy77,gm82}.
Gravitational forces are calculated using a tolerance parameter $\theta=0.7$ and including terms up to the quadrupole order in the multiple expansion. A Plummer potential is used to soften gravitational forces, with constant softening lengths for different species of particles.
In the simulations described here, if not explicitely otherwise indicated, we assume $\epsilon=280\mathrm{pc}$. \\
SPH (Smoothed particle hydrodynamics) is a Lagrangian technique in which the gas is partitioned in fluid elements represented by particles, which obey equations of motion similar to the collisionless component, but with additional terms describing pressure gradients, viscous forces and radiative effects in gas. To capture shocks, a conventional form of the artificial viscosity is used, with parameters $\alpha=0.5$ and $\beta=1.0$ \citep{hk89}.  
To describe different spatial dynamical range, SPH particles have individual smoothing lengths $h_i$, calculated in such a way that a constant number of neighbors is contained within $2h_i$. All the simulations have been performed using a number of neighbors $N_s\sim 15$.
The gas is modeled as isothermal, with a temperature $T_{gas}=10^4 K$. Because of the short cooling time of disk gas, fluctuations in the gas temperature are quickly radiated away, so that simulations employing an isothermal equation of state differ little from more realistic ones \citep{mih96,naa06}.\\
The equations of motion are integrated using a leapfrog algorithm with a fixed time step $\Delta t=5\times10^5 \mathrm{yr}$.
\subsection{Star Formation and continuous stellar mass loss}\label{sfmet}
Including star formation in gas dynamics is not a trivial task and a lot of different recipes and numerical methods can be adopted \citep{kat92,ste94,spri00,spr03,cox06}, in order to model the star formation rate on one hand, and taking into account the effects that this star formation has on the surrounding, on the other hand.\\

As in \citet{mih94b}, we parametrized the star formation efficiency for a SPH particle as
\begin{equation}\label{loc}
\frac{\dot{M}_{gas}}{M_{gas}}=C\times {\rho_{gas}}^{1/2}
\end{equation} 
with the constant $C$ chosen such that the isolated disk galaxies form stars at an average rate of between $ 1$ and $2.5 M_{\odot} yr^{-1}$. \\
The choice of the parametrization in Eq.\ref{loc} is consistent with the observational evidence that on global scales the SFR in disk galaxies is well represented by a Schmidt law of the form $\Sigma_{SFR}=A{\Sigma_{gas}}^N$, being $\Sigma_{gas}$ and $\Sigma_{SFR}$ disk-averaged surface densities, with the best fitting slope $N$ about 1.4 \citetext{see \citealp{ken98}, but also \citealp{wb02,bois03,gs04}}. Interestingly, this relation seems to apply, with a similar slope, also to local scales, as shown in \citet{ken05} for M51.\\
Once the SFR recipe is defined, we apply it to SPH particles, using the hybrid method described in \citet{mih94b}: it consists in representing each gas particle with two mass values, one refering to its gravitational mass $M_i$, whose value stays unchanged during the whole simulation, and the other describing the gas content of the particle $M_{i,gas}$, whose value changes in time, according to Eq.\ref{loc}. Gravitational forces are always evaluated on the gravitational mass $M_i$, while hydrodynamical quantities, in turn, uses the time-varying  $M_{i,gas}$.
Only if the gas fraction present into the hybrid particles drops below the $5\%$ of the initial gas content, the hybrid particle is totally converted into a star-like particle and the little amount of gas material still present is spread into the neighbors.

We also followed the method described in \citet{mih94b} for including the effects of star formation into the ISM.
The method is fully described in the above cited paper, and here we briefly recall only the main features.\
A Miller-Scalo stellar mass function is adopted, and we evaluated the fraction of stars with masses  $> 8 M_{\odot}$, assuming that they instantaneously become supernovae, living behind remnants of $1.4 M_{\odot}$ and releasing their mass on the surrounding ISM. The mass released enriches also the metallicity of the surrounding gas. This is done assuming a yield $y=M_{ret}/M_{*}$=0.02, $M_{ret}$ being the total mass of all reprocessed metals and $M_{*}$ the total mass in stars. For each gas particle, mass and metals return is applied to the $i-th$ neighbor gas particle, using a weight $w_i$ based on the smoothing kernel.  \\
The energy injection in the ISM from SNe explosions is treated assuming that only a fraction $\epsilon_{kin}$ of $E_{SN}=10^{51} \mathrm{erg}$ goes into kinetic energy, by applying a radial kick to velocities of neighbor gas particles; thus, for each SNe explosion, the  $i-th$ neighboring gas particle receives a velocity impulse directed radially away from the ``donor'', with a magnitude
\begin{equation}\label{dv}
 \Delta v_{i}=\left(\frac{2 \ w_i \ \epsilon_{kin} \ E_{SN}}{M_i}\right)^{1/2}, 
\end{equation}
$w_i$ being, once again, the weighting based on the smoothing kernel and $M_i$ the mass of the receiver.\\

Evidently, this method has a certain number of free parameters, that clearly influence not only the star formation evolution, but, more generally, the global galaxy dynamics.
Before moving on to perform simulations of galaxy encounters, we run a set of simulations of isolated galaxies, in order to check the dependence of the results on the gravitational smoothing length $\epsilon$, and on the fraction $\epsilon_{kin}$ of kinetic energy that is released, via SNe explosions, to the surrounding gas particles. For example, for the choice of the gravitational smoothing length, we performed 21 simulations of galaxies gSa, gSb and gSd, with $\epsilon=70, 140, 210, 280, 350, 420$ and $490$ pc, respectively.
The results showed that a too small value of $\epsilon$ causes a conspicuous heating of the stellar disk, for all the morphological galactic types. An $\epsilon=280 pc$, which corresponds (for gSa and gSb galaxies) to the average distance among stars in the disk seemed to represent a compromise between accuracy in the gravitational forces evaluation and not excessive relaxation effects. Also for the choice of the  $\epsilon_{kin}$ parameter, a set of numerical simulations of isolated galaxies was performed, and the results found were in good agreement to those of \citet{mih94b}, i.e. it was checked that if the total amount of kinetic energy received by a gas particle, due to the contribution from all its neighbors, is $\leq 1 \mathrm{kms^{-1}}$, this prevents the gaseous disk from a rapid growth of the vertical thickness, giving good results in terms of  gas and star formation morphology. \\

If star formation acts in consuming the gas of a galaxy, an exhaustive modelling of galactic evolution has to take into account also 
the competing process of stellar mass-loss. Indeed, as both observational studies and stellar evolutionary models propose, it can play an important role,  since the gas mass fraction restituted by stars may reach  some 45\% over the Hubble time, when integrated over the stellar mass spectrum.\\
We included continuous stellar mass loss in the model adopting the formula given in Eq.2 of \citet{jun01}, and applying it only to stellar populations formed into hybrid particles, i.e., at each time step, an amount
\begin{equation}
M_{i,s}(t)=\frac{\left(M_i-M_{i,gas}(t)\right)\Delta t\ c_0}{t-t_{birth}+T_0}
\end{equation}
of the stellar mass in the population is lost by evolutionary effects, going to enrich the gas content $M_{i,gas}$ of the hybrid particle. In the formula above, $t_{birth}$ represents the birth time of the population, $T_0=4.97 Myr$ and $c_0=5.47\times10^{-2}$ \citep[see][for details]{jun01}.

%_________________________________________________________________________

%-------RESULTS
%_________________________________________________________________________

\section{Results}\label{results}

In this Section, the main results of our study will be presented and discussed.
This first paper being devoted to the study of star formation in interacting galaxy pairs, we will exclude from our analysis 24 simulations from the total sample of 240, these simulations involving only gE0-gE0 encounters (``dry'' interactions).\\
After presenting an image gallery of galaxy interactions and discussing the main morphological features arising during encounters (Section \ref{image}), we will move to examine the evolution of the star formation.
 To distinguish which contribution to the global SFR comes from the undisturbed galaxies and which from the interaction, we will first analyze the evolution of the star formation rate for the isolated galaxies (Section \ref{sfrisol}), then we will proceed to describe the most salient features of the SFR evolution for the interacting pairs (Section \ref{sfrpairs}). The evolution of the star formation efficiency (SFE) too will be discussed (Section \ref{sfepairs}), while in Section \ref{where} we will describe where star formation regions are located.  \\
The next step will consist in trying to understand which parameters determine the different SFR obtained soon after the pericenter passage and in the merging phase, for different galaxy pairs. This will lead us to study in detail the dependence of the SFR on: total gas content $M_{gas}$ in the galaxy just before the burst phase (Section \ref{does1}), galaxies separation at pericenter passage $R_p$, galaxies relative velocity at pericenter passage $V_p$, characteristic time of the encounter $t_{enc}$, tidal effects of the encounter, quantified by means of a suitable 'tidal parameter' (Section \ref{does2}). The discussion on the dependence of the SFR on this tidal parameter will be deepened in Section \ref{tidal}, while in Section \ref{toward} we will propose a formulation for the SFE in the burst phase. Section \ref{globschmidt} presents a discussion of the evolution of interacting, merging and post-merger galaxies into the $(\Sigma_{gas}, \Sigma_{SFR})$ plane.
Finally in Section \ref{gasinflow} gas inflow into the central galactic regions will be exploited.\\
 
In the following, we will often refer to galaxy center in our analysis.
If not otherwise written, we refer to a density-weighted center evaluated in the following way:\\
for each configuration, and for each galaxy,
\begin{itemize}
\item first we evaluate the density center \citep{cashut85} of the dark matter particles $C_{DM}$ initially (t=0) belonging to the galaxy;
\item then we evaluate the density center for all the particles (gas+stars+dark matter) initially (t=0) belonging to the galaxy, that are at a distance $r<10$ kpc from  $C_{DM}$;
%\item 
the density center so found is used to define the center $C$ of the whole galaxy.
\end{itemize}
\begin{figure*}
  \centering
  \includegraphics[width=9.8cm,angle=270]{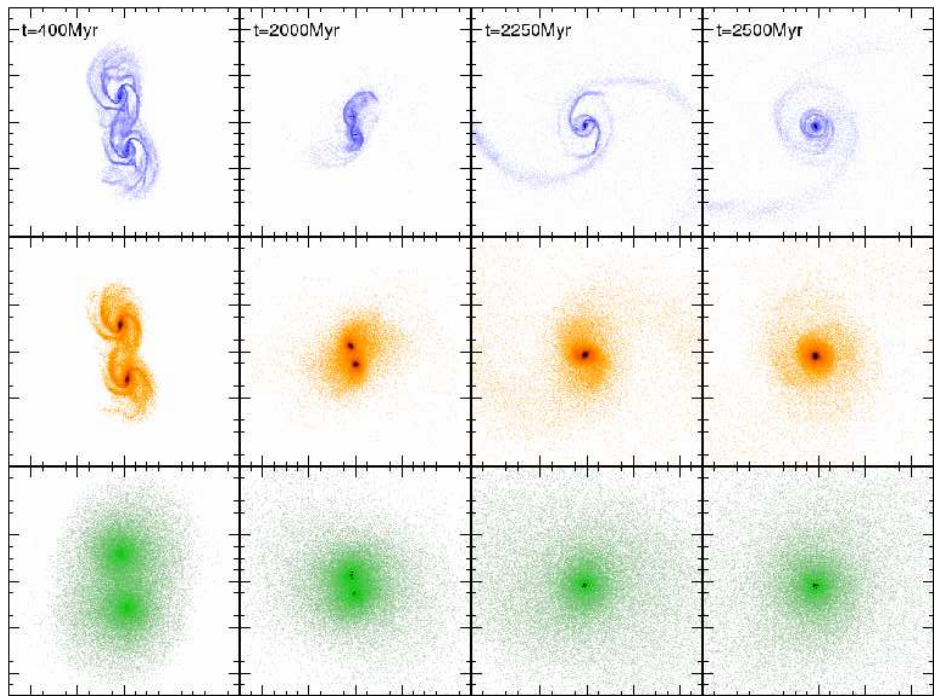}
  \caption{Evolution of gas (upper panels), stars (intermediate panels) and dark matter (lower panels) during a direct merger between two gSb galaxies (id=09dir in Table \ref{orbpos}). Time is labelled in the upper part of the Figure. Each frame is 50 kpc $\times$ 50 kpc in size.}
  \label{gSbgSb09dir00}%
\end{figure*}

\begin{figure*}
  \centering
  \includegraphics[width=9.8cm,angle=270]{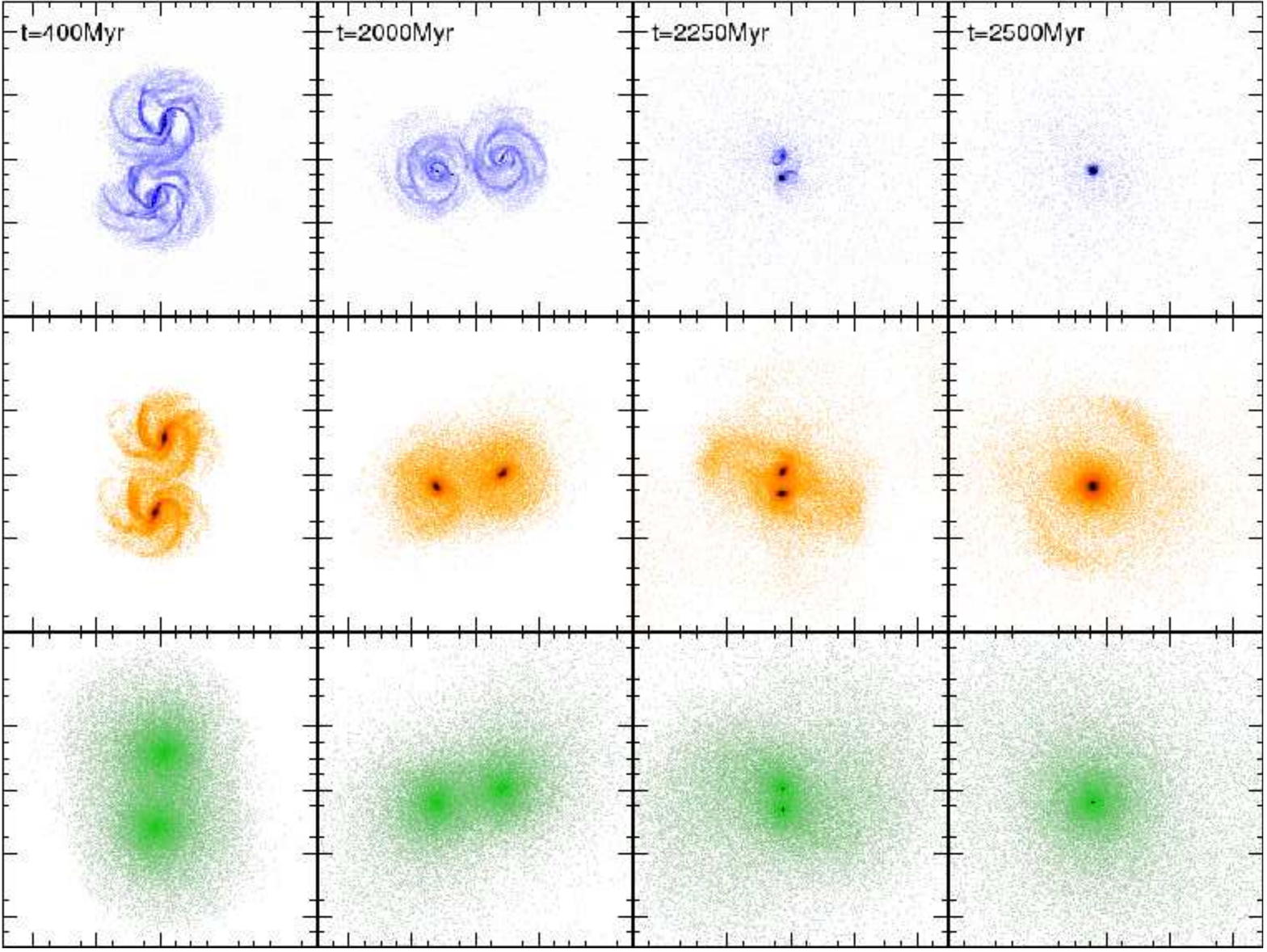}
  \caption{Evolution of gas (upper panels), stars (intermediate panels) and dark matter (lower panels) during a retrograde merger between two gSb galaxies (id=09ret in Table \ref{orbpos}). Time is labelled in the upper part of the Figure. Each frame is 50 kpc $\times$ 50 kpc in size.}
  \label{gSbgSb09ret00}%
\end{figure*}

\begin{figure*}
  \centering
  \includegraphics[width=9.8cm,angle=270]{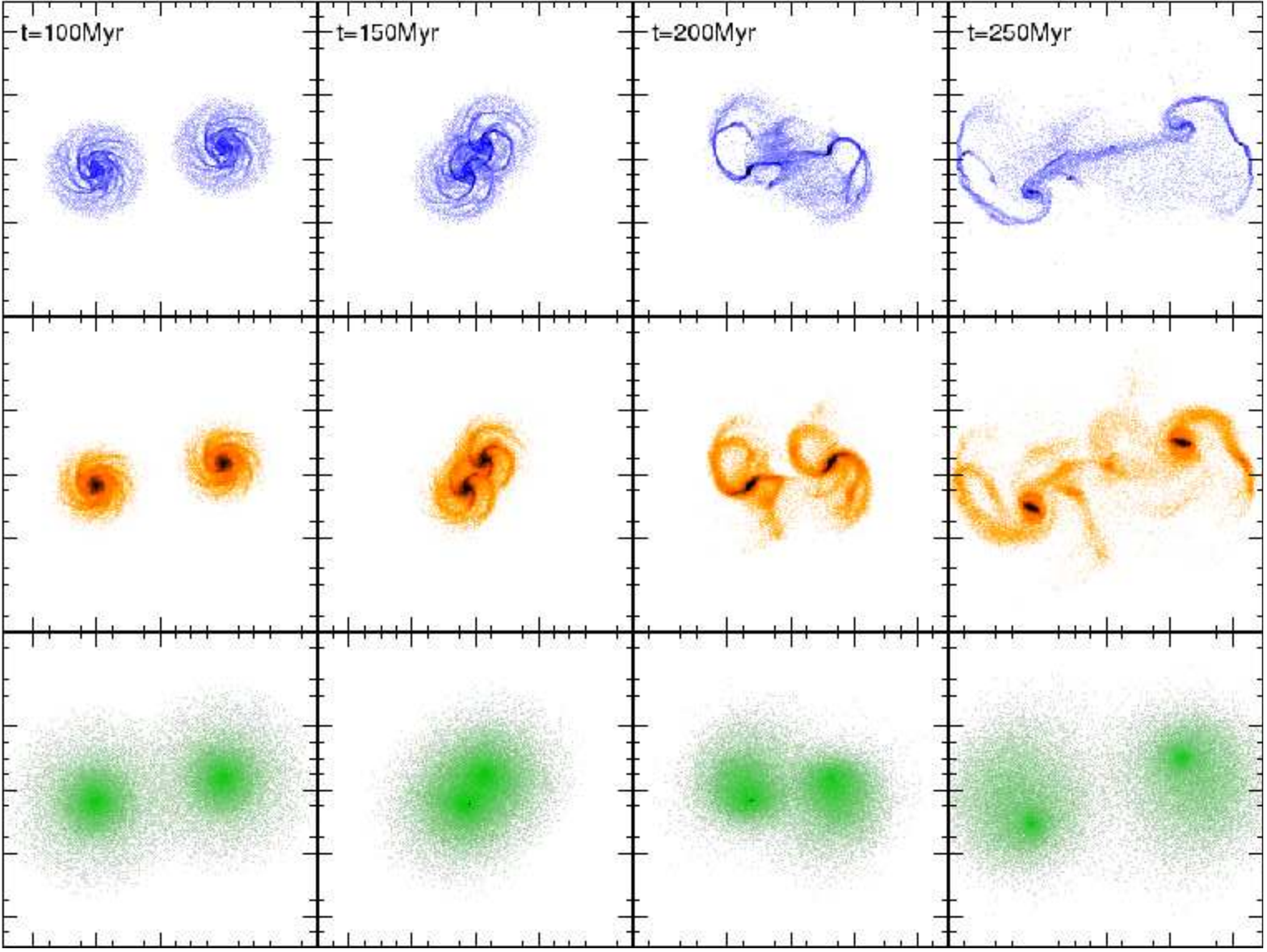}
  \caption{Evolution of gas (upper panels), stars (intermediate panels) and dark matter (lower panels) during a direct flyby between two gSa galaxies (id=04dir in Table \ref{orbpos}). Time is labelled in the upper part of the Figure. Each frame is 50 kpc $\times$ 50 kpc in size.}
  \label{gSagSa04dir00}%
\end{figure*}
%%%%%

For evaluating the galaxy velocity, we evaluated the mass-weighted velocity of all the particles (stars+gas+dm) at a distance $r<10 \mathrm{kpc}$ from $C$.\\
We will also refer sometimes to specific encounters. In this case the nomenclature adopted is the following: morphological type of the two galaxies in the interaction (gE0, gSa, gSb or gSd), + the encounter identification string (see first column in Table\ref{orbpos}), + 00 (for indicating a planar encounter). For example the nomenclature gE0gSa04ret00 corresponds to a planar interaction between an elliptical and a Sa spiral, whose initial orbital parameters are those corresponding to id=04ret in Table\ref{orbpos}.

\subsection{A gallery of galaxy interactions}\label{image}
%%%%the following figures are discussed in ``A gallery of galaxy interactions''

Given the large number of simulations performed, it is not possible to describe each case individually. So we will proceed to describe some 'fiducial' cases, that retain the mean features of all the simulations performed. In Figs.\ref{gSbgSb09dir00}, \ref{gSbgSb09ret00} and \ref{gSagSa04dir00}, some sequences of galaxy mergings and flybys  are shown.\\

Fig. \ref{gSbgSb09dir00} describes the encounter and successive merger of two gSb galaxies, on  direct orbit. At the beginning of the simulation the two galaxies are separated by a distance of 100 kpc. As they start to approach each other, they begin to develop tails, populated of both stars and gas particles. The intense tidal field exerted during the pericenter passage (t=400 Myr) leads also to a transfer of mass between the two systems. 
In many cases, as it will be shown in the next sections, the direct encounter leads to a more rapid and dramatic merger, with an expansion of the outer parts of the system, which is particularly visible for disk galaxies.\\

\begin{figure}
  \centering
  \includegraphics[width=5.cm,angle=270]{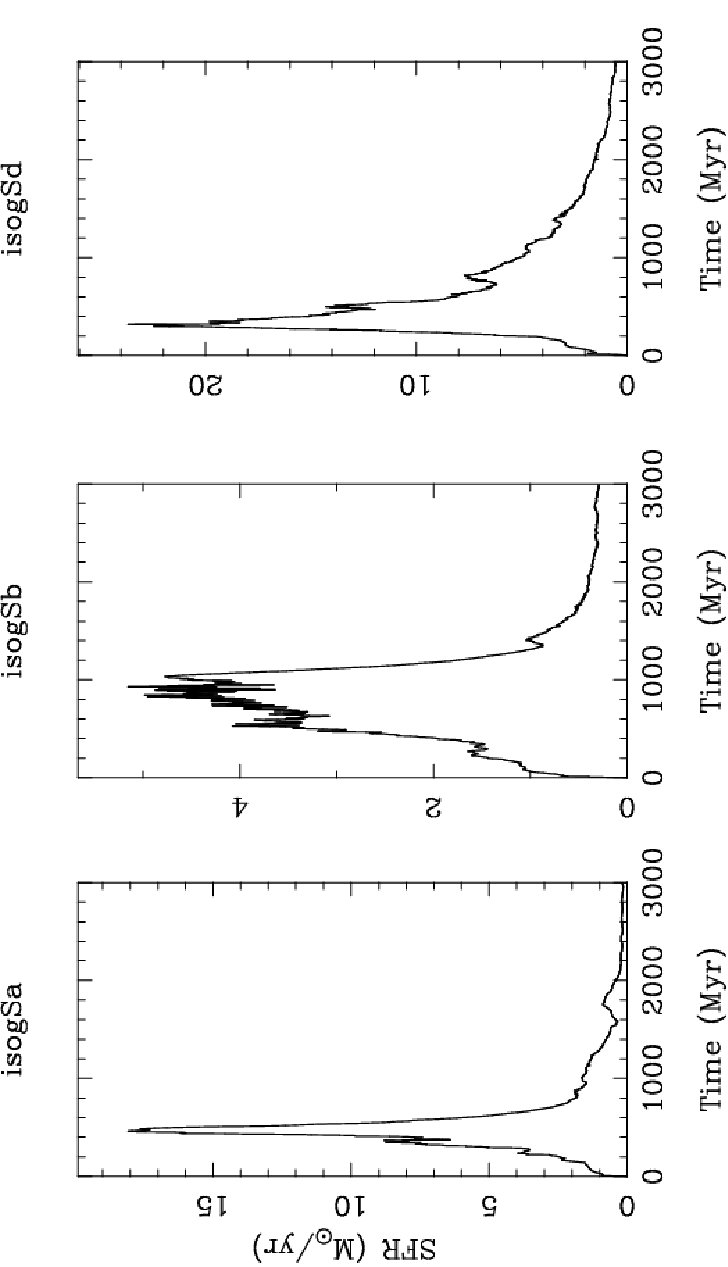}
  \caption{From left to right: Star formation rate of the isolated gSa, gSb and gSd galaxies.}
  
  \label{sfriso}%
\end{figure}

\begin{figure}
  \centering
  \includegraphics[width=4.cm,angle=270]{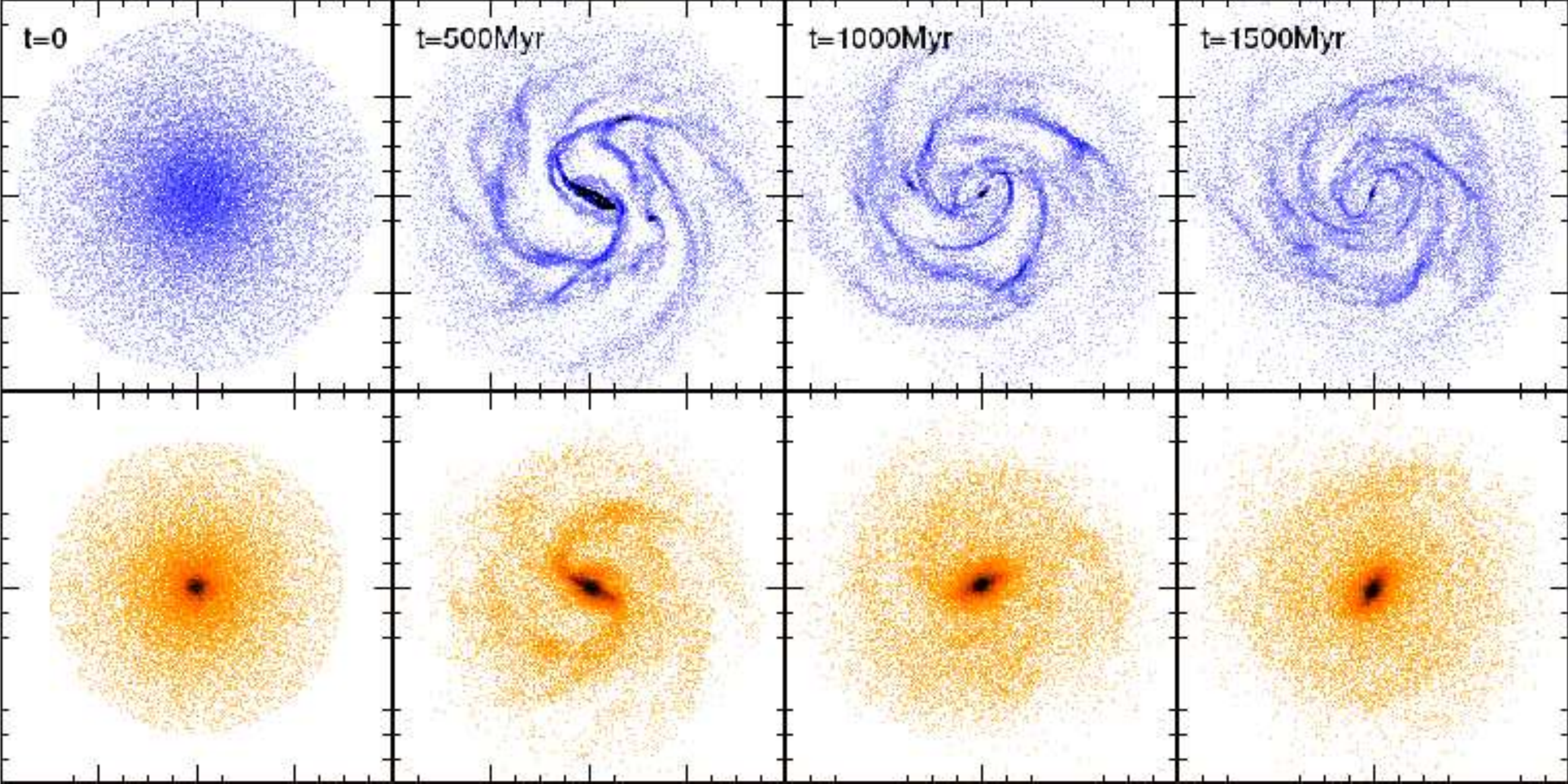}
  \caption{Evolution of the gas component (upper panels) and of the stellar one (lower panels) for the isolated gSb galaxy.  Time is labelled in the upper part of the Figure. Each frame is 20 kpc $\times$ 20 kpc in size.}
  
  \label{tot}%
\end{figure}

The retrograde merger is shown in Fig. \ref{gSbgSb09ret00}. In this case the two galaxies are less affected by the tidal interaction. The disk galaxies develop two transient great spiral arms, after the pericenter passage, but retain much longer their initial structure. Contrary to the direct case, here no transfer of mass from one galaxy to the other takes place and the formation of tidal tails is less obvious. As it will be discussed later, this allows most of the initial gas mass to stay well confined in the disk of the spiral galaxy, furnishing a great reservoir for the intense starburst that takes place in the final stage of the merging. Note indeed the high gas concentration in the inner central galactic regions in the last two snapshots of galaxies in Fig.\ref{gSbgSb09ret00}.\\
Finally, Fig. \ref{gSagSa04dir00} shows a direct flyby between two gSa spirals. Tidal tails and a bridge connecting the two galaxy
 centers develop after the pericenter passage at t=170Myr. In this, as in the previous Figures, the high density regions in the gas component correspond also to the sites of most intense star formation, as it will be discussed in Section \ref{where}.

\subsection{Evolution of  isolated galaxies}\label{sfrisol}

Since the goal of this paper is to investigate the role interactions and mergers play in the star formation process, it is essential to study the evolution of the SFR in the isolated galaxies, in order to distinguish secular evolution from tidal effects in the interacting galaxy sample. This is shown in Fig. \ref{sfriso}. In all the three cases, an initial burst develops, driven by the compression of the gas into density waves.  In Fig.\ref{tot}, which shows the time evolution of the gaseous and stellar component for the gSb galaxy, spiral arms and a bar are clearly formed at t=500Myr.\\
%%%
%%%%%THe 2 following fig are discussed in the next section
\begin{figure}
  \centering
  \includegraphics[width=13cm,angle=270]{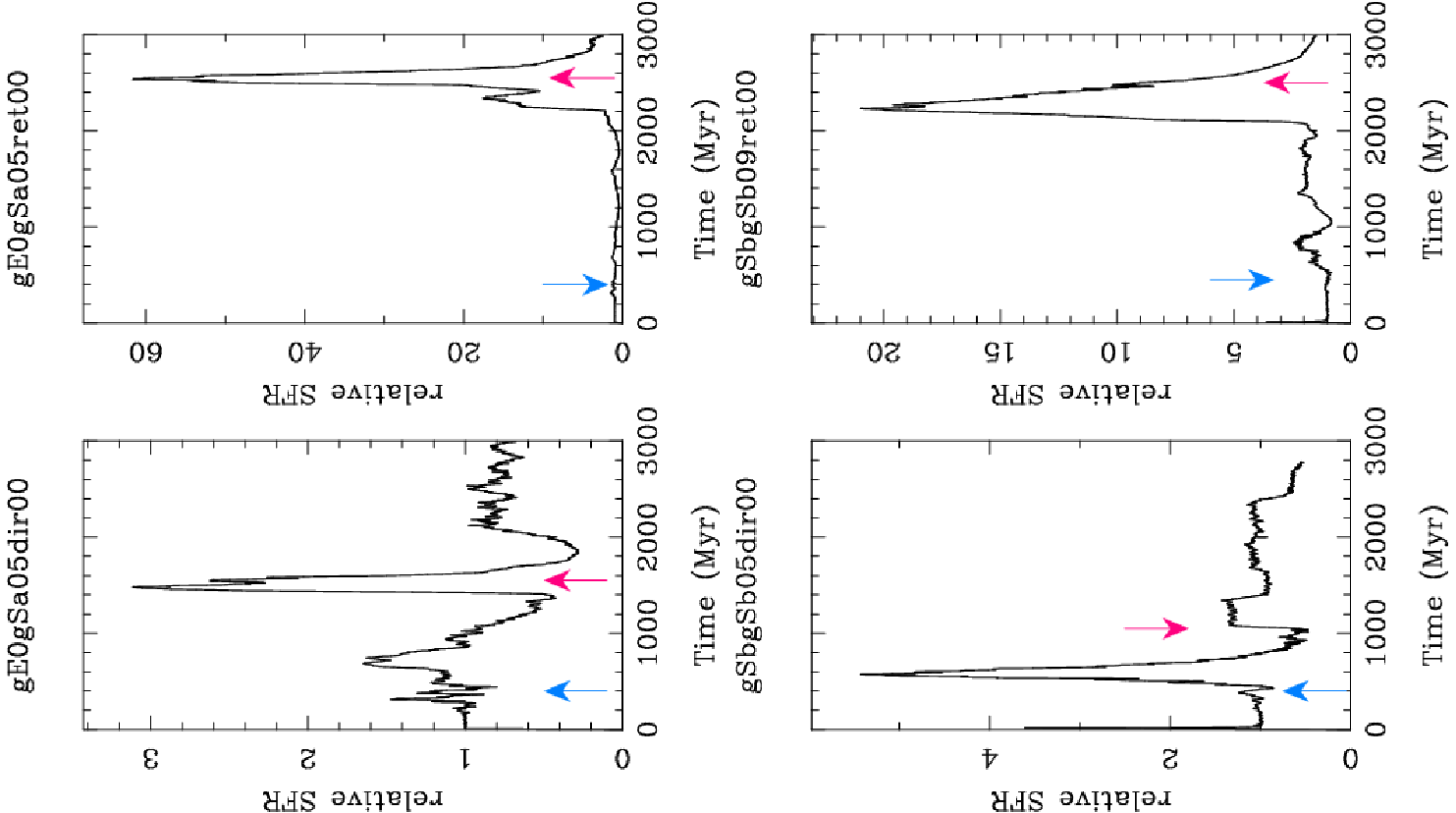}
  \caption{Star formation rate, versus time, for some galaxy mergers. The SFR is normalized to that of the corresponding isolated galaxies. The blue arrows indicate the first pericenter passage between the two galaxies, and the red arrows the merger epoch.}
  
  \label{sfrint}%
\end{figure}

\begin{figure}
  \centering
  \includegraphics[width=13cm,angle=270]{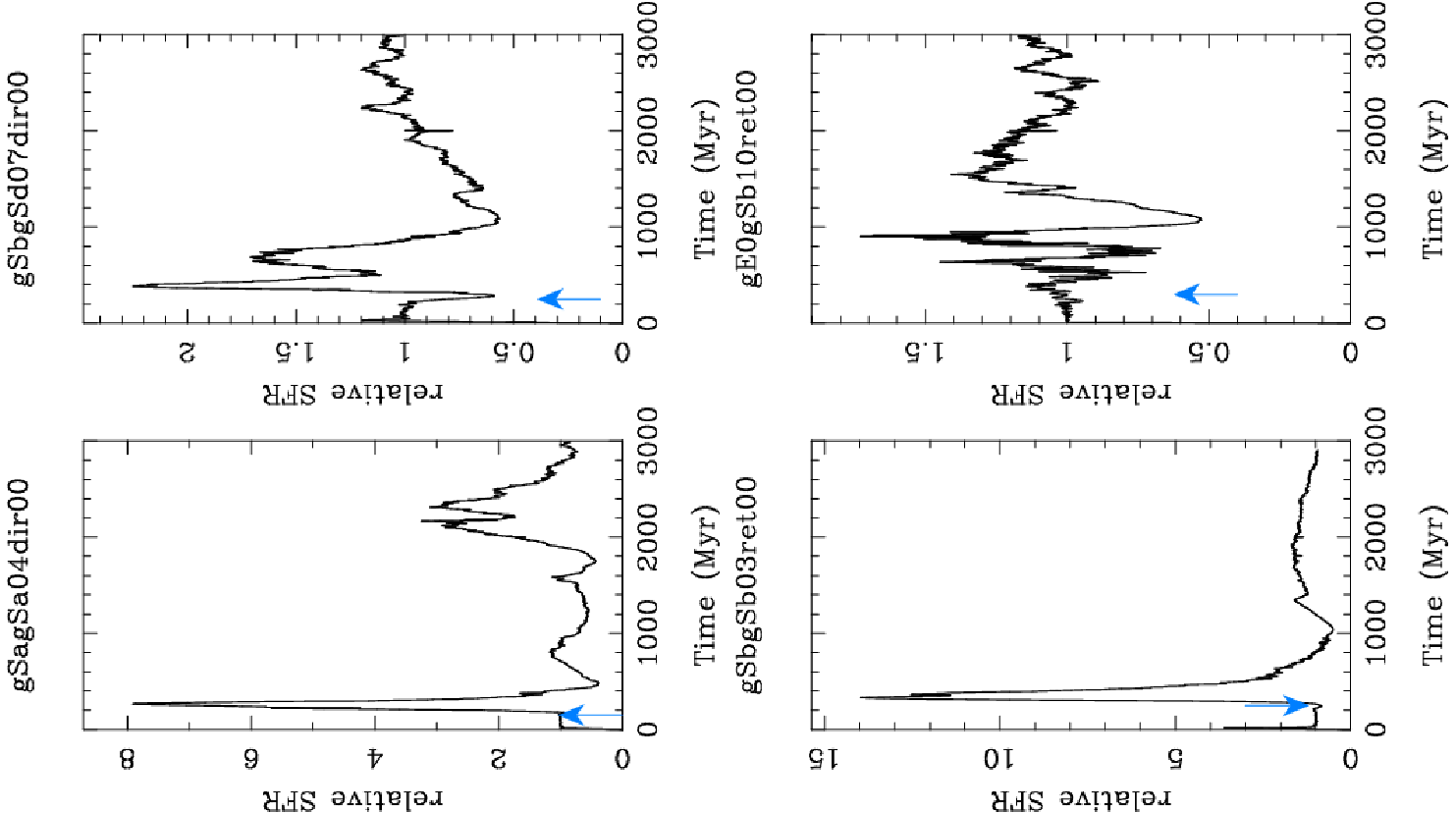}
  \caption{Star formation rate, versus time, for some galaxy flybys. The SFR is normalized to that of the corresponding isolated galaxies. The blue arrows indicate the epoch of pericenter passage between the two galaxies.}
  
  \label{sfrfly}%
\end{figure}
%%%%%%%%%%%%%

After the initial peak, the SFR follows a typical exponential profile. This is a general result of simulations of  galaxies evolved in isolation \citep[see, for example, the discussions in][]{com04, har06}, unless \emph{ad hoc} recipes for modeling feedback from SNe explosions are taken into account \citep{spri00}. To reproduce the average constant star formation rates found in observations of spiral galaxies in the middle of the Hubble sequence \citep{ken83, ken94}, it is indeed necessary to take into account also external gas accretion, as shown in simulations of galaxy evolution in a cosmological frame \citep{tis00, nag04}.\\
Averaged on the first $3$ Gyr of evolution, the $<$SFR$>$ varies from $1 \mathrm{M_{\odot}/yr}$ to $2.5 \mathrm{M_{\odot}/yr}$.

\subsection{Star formation in interacting pairs}\label{sfrpairs}

\begin{figure}
  \centering
  \includegraphics[width=9.cm,angle=270]{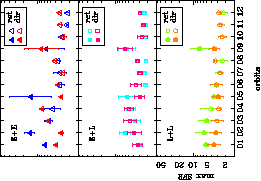}
  \caption{Maximum star formation rate, relative to the isolated case, as a function of the orbital type, for flybys (empty symbols) and mergers (solid symbols). Encounters are grouped into three classes, depending on the morphology of the interacting galaxies: interactions between two early type galaxies are shown in the upper panel, the medium panel refers to early-late type encounters, the lowest to encounters involving two late-type systems. In each panel, error bars represent the standard error of the mean. The large error bars found in some cases reflect a large dispersion in the data.}
  
  \label{max-orb}%
\end{figure}

%%%%the following figure is discussed in the next Section
%-----------------------------------------------------------------
\begin{figure}
  \centering
  \includegraphics[width=9.cm,angle=270]{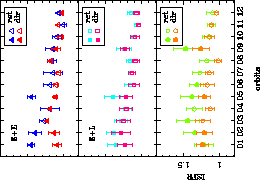}
  \caption{Integrated star formation rate, ISFR, relative to the isolated case, as a function of the orbital type, for flybys (empty symbols) and mergers (solid symbols). Encounters are grouped into three classes, depending on the morphology of the interacting galaxies: interactions between two early type galaxies are shown in the upper panel, the medium panel refers to early-late type encounters, the lowest to encounters involving two late-type systems. In each panel, error bars represent the standard error of the mean. The large error bars found in some cases reflect a large dispersion in the data.}
  \label{pisfr-orbgtutte}%
\end{figure}
%-----------------------------------------------------------------

The variety of morphological and orbital parameters adopted in the simulations reflects into the different star formation histories of galaxies in the sample. Some SFR evolution with time are shown in Figs.\ref{sfrint} and \ref{sfrfly}, which refer, respectively, to mergings and flybys. Simply looking at these few examples, it appears clear that interactions can increase star formation in the galaxy pairs from low levels (1.5-2 times the isolated case) to values typical of starburts galaxies (20-60 times the isolated case).\\
In these figures, no distinction is made between the two interacting galaxies in the sample, i.e. the relative SFR at time $t$ is evaluated as the ratio between the SFR of the pair and the total SFR of the isolated galaxies, both evaluated at time $t$.     
From Fig.\ref{sfrint}, one can extrapolate some of the main features of the star formation evolution for interactions leading to mergers:
\begin{itemize}
\item  the peak of the SFR usually occurs in the last phases of the encounter, when the two galaxies are interpenetrating, while, at the first pericenter passage, the increase in the SFR is only modest (1 to 4-5 times that of the isolated case).
\item for the same orbit, direct and retrograde encounters can lead to different SFR evolution. Obviously, this is not a general rule, but often the SFR is greater for retrograde mergers than direct ones and in some cases the differences between direct and retrograde encounters is conspicuous. For example in the gE0gSa05ret00 encounter, the maximum SFR is 20 times greater than of the direct gE0gSa05dir00 encounter (upper panels in the Figure). 
\end{itemize}  

For flybys, the peak in SFR occurs usually just after the pericenter passage; also in this case the increase in the star formation rate can be modest (as for the gSbgSd07dir00 and the  gE0gSb10ret00 cases in Fig.\ref{sfrfly}) or substantial (cf. the gSagSa04dir00 and gSbgSb03ret00 in the same Figure). On average, the SFR increases by a factor 3.4 with respect to the isolated galaxy sample.\\

Note that, both in Fig.\ref{sfrint} and Fig.\ref{sfrfly}, the peak in the star formation rate corresponds to the occurence of a nuclear starburst.\\

To show the wide variety of SFR obtained, in Fig. \ref{max-orb} we report the maximum SFR obtained during the 3 Gyr-evolution of the pairs, as a function of the different orbits simulated. As before, the SFR is relative to the isolated case. In this plot, results are shown for the whole sample (i.e. both mergers and flybys). It clearly puts in evidence that, for the same initial orbital conditions, different morphologies of galaxies in the pair lead to different SFR histories, but also that, fixing the morphology of the pair, direct and retrograde encounters usually do not lead to the same star formation evolution. For mergers in general \emph{retrograde encounters are more efficient that direct ones in driving star formation}: indeed, on average, in retrograde mergers the SFR peaks to 8.9 times that of the corresponding isolated galaxies, while, for direct mergers, the maximum star formation rate (always relative to the isolated case) is only 5.

\subsubsection{Integrated star formation}\label{isfr}

%%%the following two figures are discussed in the next Section
%-----------------------------------------------------------------
\begin{figure}
  \centering
  \includegraphics[width=13cm,angle=270]{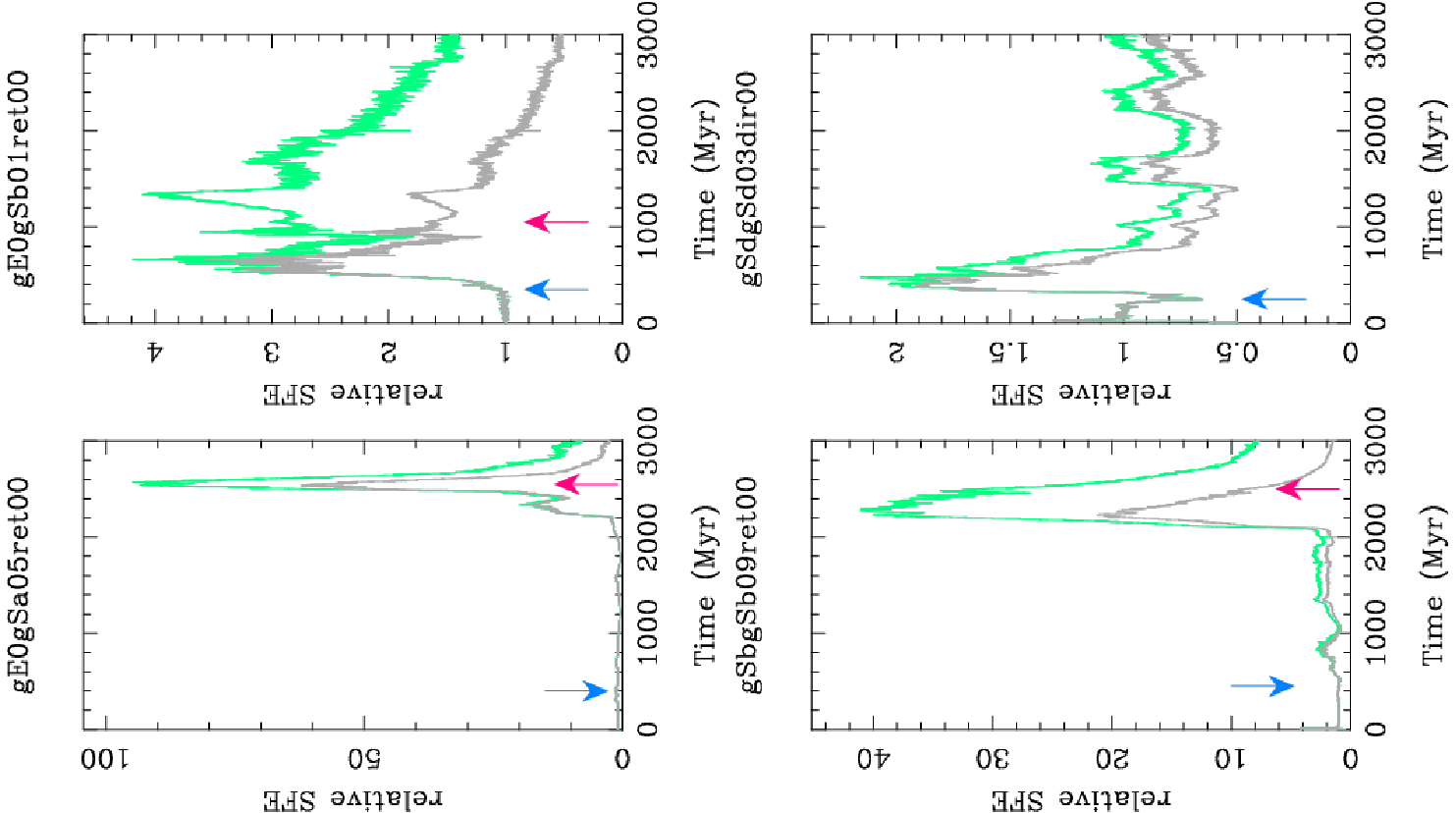}
  \caption{Star formation efficiency (cyan curve), relative to the isolated case, as a function of time for three mergers (gE0gSa05ret00, gE0gSb01ret00, gSbgSb09ret00) and a flyby (gSdgSd03dir00). The blue arrows indicate the first pericenter passage between the two galaxies, and the red arrows the merger epoch. For comparison also the relative SFR is shown (grey curve)}
  \label{psfe}%
\end{figure}

\begin{figure}
  \centering
  \includegraphics[width=9.cm,angle=270]{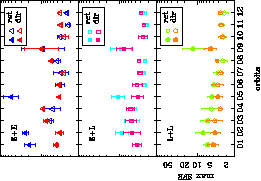}

  \caption{Maximum star formation efficiency, relative to the isolated case, as a function of the orbital type, for flybys (empty symbols) and mergers (solid symbols).  Encounters are grouped into three classes, depending on the morphology of the interacting galaxies: interactions between two early type galaxies are shown in the upper panel, the medium panel refers to early-late type encounters, the lowest to encounters involving two late-type systems. In each panel, error bars represent the standard error of the mean. The large error bars found in some cases reflect a large dispersion in the data.} 
  \label{maxsfe-orb}%
\end{figure}
%-----------------------------------------------------------------
Here we want to examine the evolution of the integrated star formation rate $\left(ISFR=\int_{t=0}^{t=3 Gyr}{SFR(t)dt}\right)$, i.e. the total quantity of gas transformed into stars, during the whole duration of the simulations.
We have seen that different simulations show different SFRs, therefore they are more or less efficient, during pericenter passages or mergers, in transforming gas into stellar matter. It is interesting to check if the ISFR shows the same trend (for example if retrograde mergers are more efficient than direct ones), because a priori this is not assured: ISFR can be dominated by an extreme starburst-like event for a short duration (several million years) or a long-term low enhancement of SFR due to interaction.\\
In Fig.\ref{pisfr-orbgtutte}, the ISFR, normalized to the isolated case, is shown, for all the simulations performed, as a function of the orbit. The main features are:
\begin{itemize}
\item Integrated on 3 Gyr of evolution, the total gas mass converted into stars can be increased by up to a factor $\sim$ 2, due to galaxy interactions. 
\item Flybys can be as efficient as mergers in transforming gas into stars: for example, two gSb galaxies, involved in a retrograde flyby, can have an ISFR comparable to that obtained in a merger (cfr. the L-L flyby with orbit id=03 with the L-L mergers 05 and 09 in Fig. \ref{pisfr-orbgtutte}).
\item In general, retrograde encounters are slighlty more efficient in transforming gas into stars: on average, for retrograde interactions $<ISFR>=1.24$, while $<ISFR>=1.15$ for direct encounters.
\item The highest ISFR values are obtained for interactions involving late type galaxies.
\end{itemize}

\subsection{Star formation efficiency in interacting pairs}\label{sfepairs}

To describe how much stars are produced by unit gas mass, it is useful to study the star formation efficiency (SFE).
This parameter can be defined as the ratio between the amount of gas transformed into stars, at a generic time $t$, and the available gas content at the same time:
\begin{equation}\label{sfe}
SFE(t)=\frac{M_{gas\to \star}(t)}{M_{gas}(t)}
\end{equation} 

Several authors indicate that strongly interacting galaxies are also more efficient in forming stars \citep{sol88,com94,geo00}, while others \citep{casa04} found that even if interacting galaxies appear more luminous in the infrared, this higher star formation rate does not correspond to a different efficiency in star formation per unit of calculated $H_2$ mass.

In Fig.\ref{psfe} the SFE, relative to that of the corresponding isolated galaxies, is shown for some of the simulation performed.
In all the cases, we found that an increase in the SFR is correlated with an increase  in the SFE, as expected from Eq.\ref{sfe}.
It is interesting to note, however, that, for the most intense bursts:
\begin{itemize}
\item usually the SFE peak is higher than the corresponding SFR and is also slightly delayed in time
\item the characteristic SFE times are greater than the corresponding SFR ones, i.e., after the encounter/merger many galaxies can show a SFR which is returned at preinteraction levels, while the SFE still shows higher values in comparison to the isolated counterparts.
\end{itemize}
%%%the following four figures are discussed in the next Section
%-----------------------------------------------------------------
\begin{figure}
  \centering
  \includegraphics[width=6.5cm,angle=270]{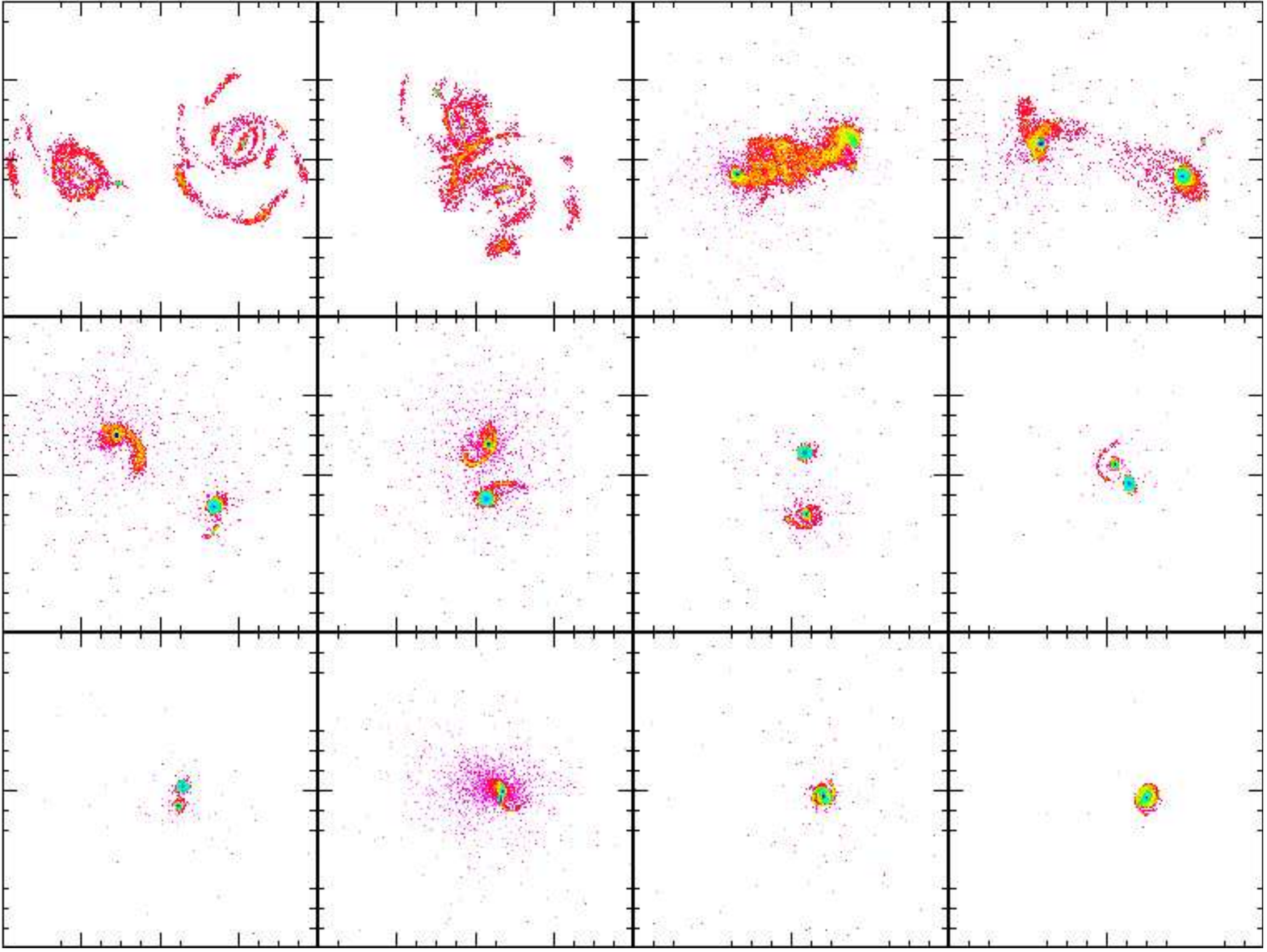}
  \caption{Star forming regions during the final stage of an encounter involving two gSb galaxies in retrograde orbits (id=09ret in Table \ref{orbpos}). %Only hybrid particles with a stellar mass fraction greater than the 30\% of the initial (t=0) particle gas mass are shown.
Only hybrid particles with a star formation efficiency (averaged on 50 Myr) greater than 0.005 are shown.
Each frame is 20 kpc in length. Snapshots are shown every 50 Myr. Blue-green colors correspond to the highest densities regions. See Fig.\ref{gasmergSbgSb09ret00} for a comparison with gas maps. }
  \label{nwsmergSbgSb09ret00}%
\end{figure}

\begin{figure}
  \centering
  \includegraphics[width=6.5cm,angle=270]{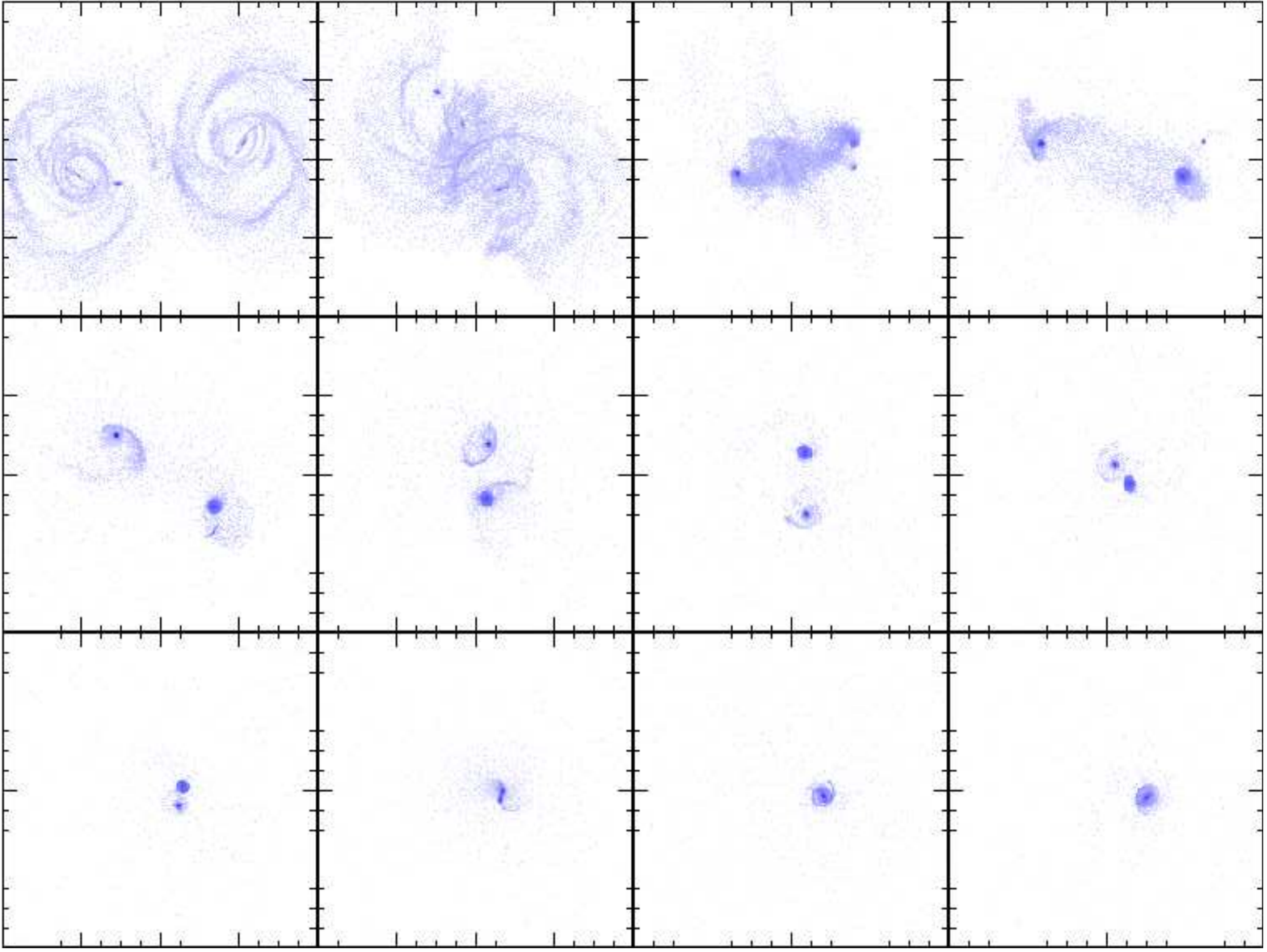}
  \caption{Gas maps during the final stage of an encounter involving two gSb galaxies in retrograde orbits  (id=09ret in Table \ref{orbpos}).
Each frame is 20 kpc in length. Snapshots are shown every 50 Myr. See Fig.\ref{nwsmergSbgSb09ret00} for a comparison with star forming regions.}
  \label{gasmergSbgSb09ret00}%
\end{figure}

\begin{figure}
  \centering
  \includegraphics[width=6.5cm,angle=270]{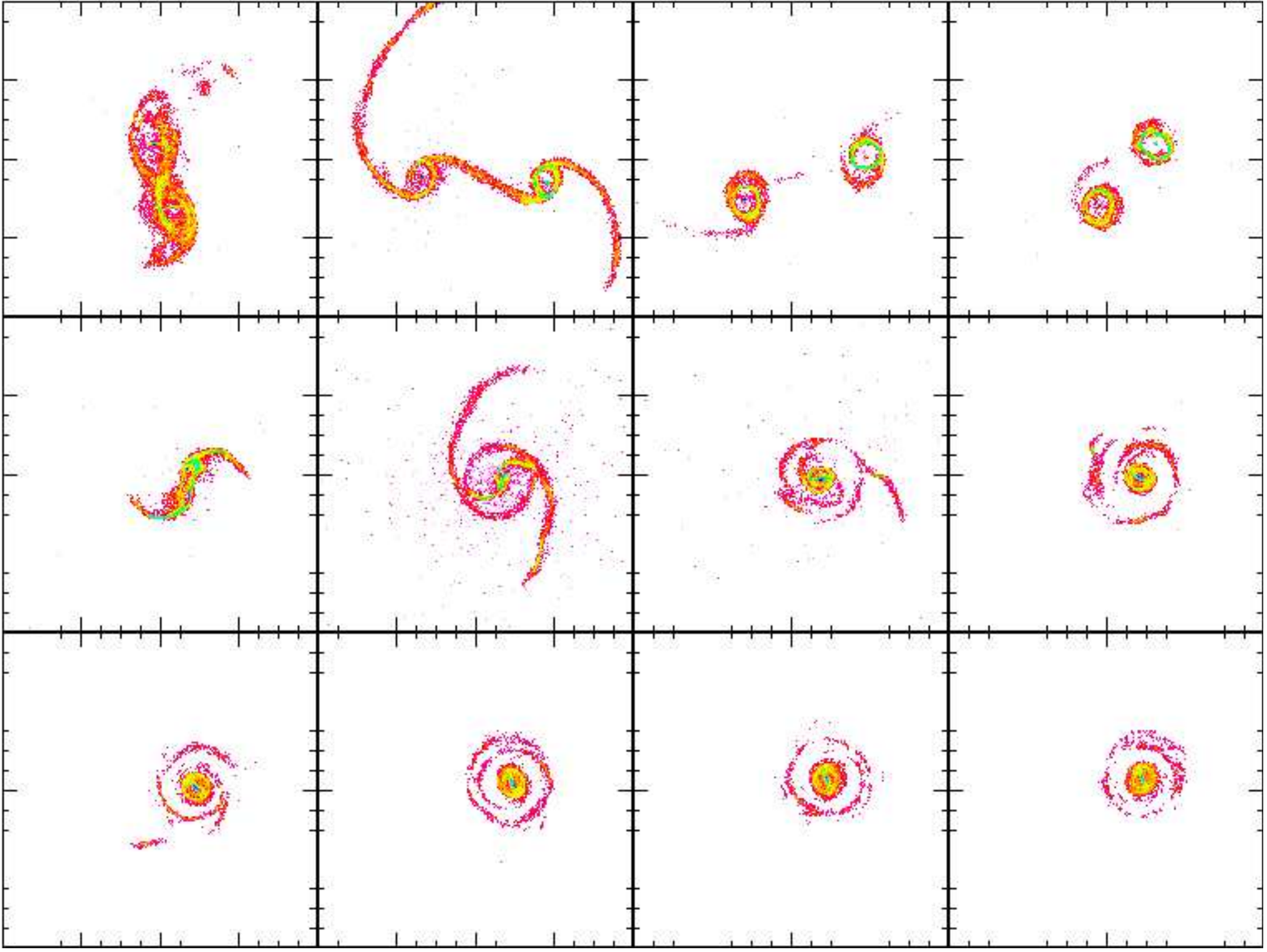}
  \caption{Star forming regions during the final stage of an encounter involving two gSb galaxies in direct orbits (id=09dir in Table \ref{orbpos}). Only hybrid particles with a star formation efficiency (averaged on 50 Myr) greater than 0.005 are shown.
%Only hybrid particles with a stellar mass fraction greater than the 30\% of the initial (t=0) particle gas mass are shown.
Each frame is 20 kpc in length. Snapshots are shown every 50 Myr.  Blue-green colors correspond to the highest densities regions. See Fig.\ref{gasmergSbgSb09dir00} for a comparison with gas maps. }
  \label{nwsmergSbgSb09dir00}%
\end{figure}

\begin{figure}
  \centering
  \includegraphics[width=6.5cm,angle=270]{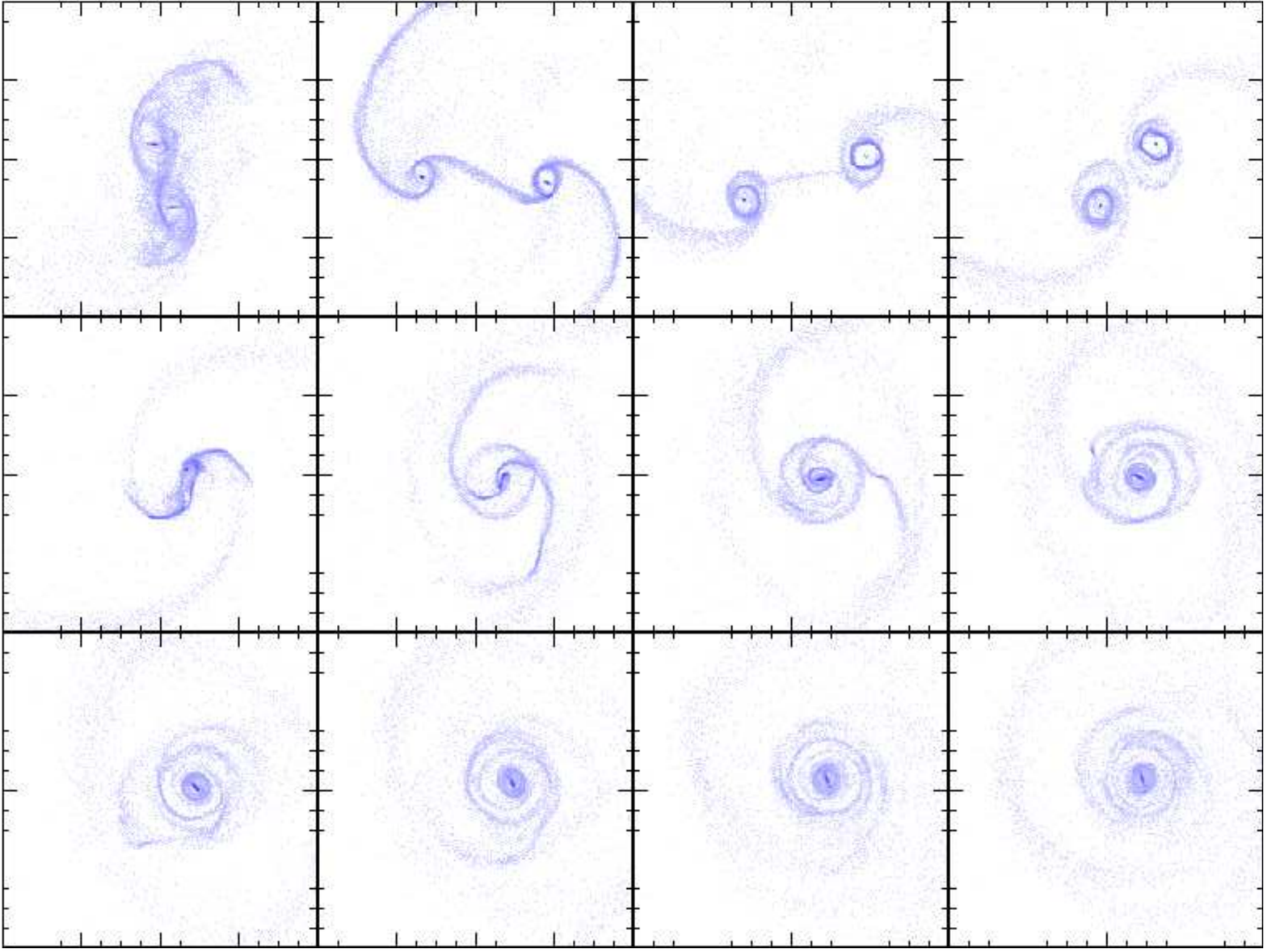}
  \caption{Gas maps during the final stage of an encounter involving two gSb galaxies in direct orbits (id=09dir in Table \ref{orbpos}).
Each frame is 20 kpc in length. Snapshots are shown every 50 Myr. See Fig.\ref{nwsmergSbgSb09dir00} for a comparison with star forming regions.}
  \label{gasmergSbgSb09dir00}%
\end{figure}
%-----------------------------------------------------------------

%%%the following  two figures are discussed in the next Section
%-----------------------------------------------------------------

\begin{figure}
  \centering
  \includegraphics[width=6.5cm,angle=270]{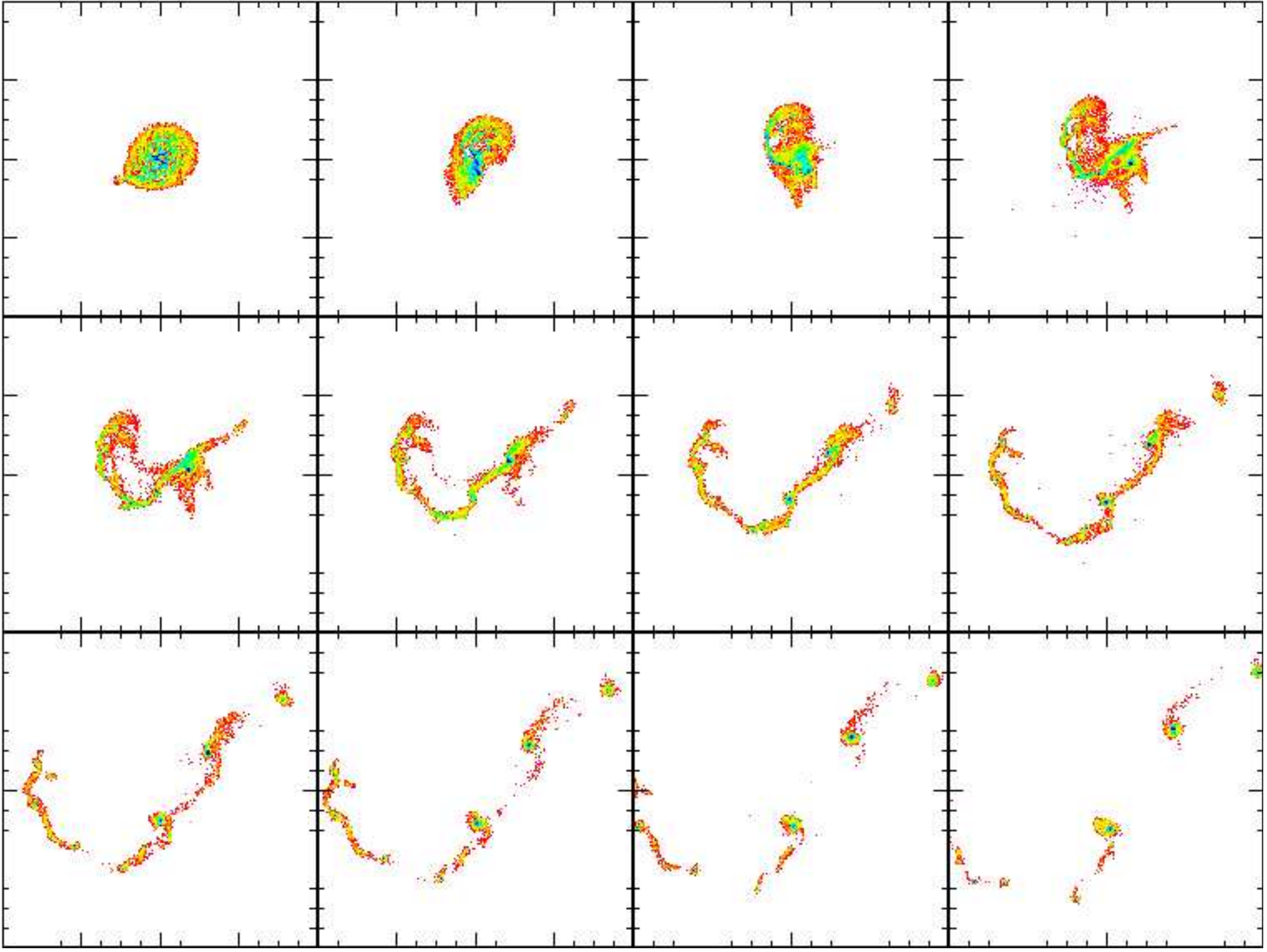}
  \caption{Star forming regions during the final stage of an encounter involving a gEO and a gSd galaxy in direct orbits (id=03dir in Table \ref{orbpos}). %Only hybrid particles with a stellar mass fraction greater than the 10\% of the initial (t=0) particle gas mass are shown.
Only hybrid particles with a star formation efficiency (averaged on 50 Myr) greater than 0.005 are shown.
Each frame is 100 kpc in length. Snapshots are shown every 50 Myr.  Blue-green colors correspond to the highest densities regions.
See Fig.\ref{gasmergE0gSd03dir00} for a comparison with gas maps. }
  \label{nwsmergE0gSd03dir00}%
\end{figure}
\begin{figure}
  \centering
  \includegraphics[width=6.5cm,angle=270]{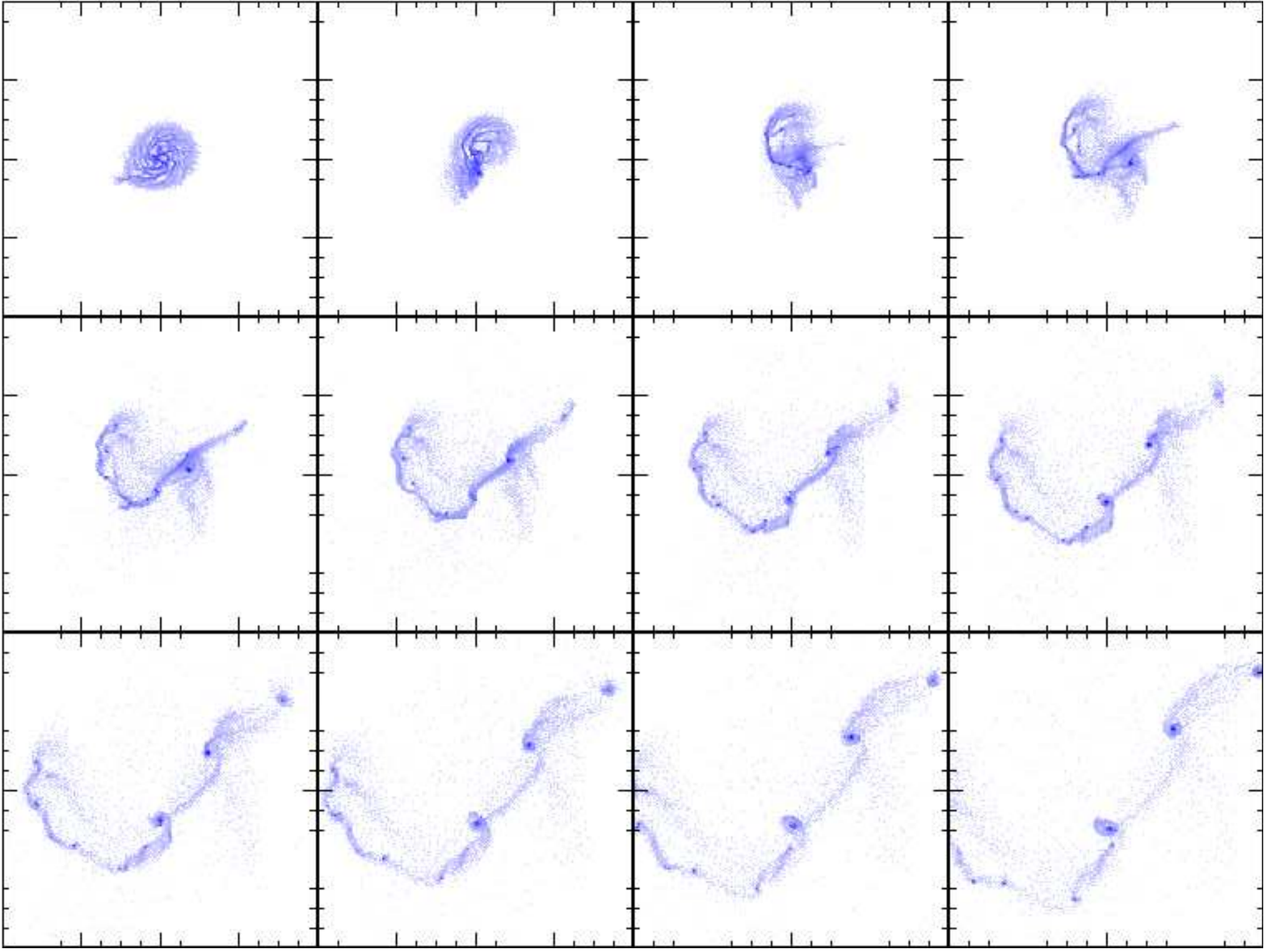}
  \caption{Gas maps during the final stage of an encounter involving a gE0 and a gSd galaxy in direct orbits (id=03dir in Table \ref{orbpos}).
Each frame is 100 kpc in length. Snapshots are shown every 50 Myr. See Fig.\ref{nwsmergE0gSd03dir00} for a comparison with star forming regions.}
  \label{gasmergE0gSd03dir00}%
\end{figure}
%-----------------------------------------------------------------

 This is evidently a consequence of the adopted SFE definition, because of the fact that, during an intense burst of star formation, gas in interacting galaxies is more rapidly depleted, with respect to the isolated case, thus the total amount of gas content in the galaxies is then lower than that of the isolated counterpart.\\

In Fig. \ref{maxsfe-orb}, the maximum star formation efficiency obtained for each encounter is shown, as a function of the orbit.

\subsection{Where do stars form?}\label{where}

%%%the following  figure is discussed in the next Section
%-----------------------------------------------------------------
\begin{figure}
  \centering
  \includegraphics[width=5.cm,angle=270]{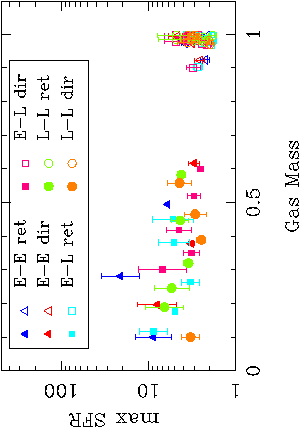}
  \caption{Maximum star formation rate (relative to the isolated case), as a function of the gas content (rescaled to its initial value), for flybys (empty symbols) and mergers (solid symbols). Different symbols correspond to different morphologies of interacting galaxies, as explained in the legend. Error bars represent the standard error of the mean. }
  \label{psfrmax-gas}%
\end{figure}

%-----------------------------------------------------------------

From the results presented in the previous sections, it is clear that interactions and mergers can lead to a conspicuous enhancement both in the SFR and in SFE (up to a factor of 100 for the SFE relative to the isolated case), but this is not always so.
The picture that begins to emerge is that interactions, and mergers in particular, can lead to intense starburst but this is not always the case, i.e. \emph{galaxy interactions are not a sufficient condition to convert high gas mass quantities into new stars} \citetext{see \citealp{berg03}  for an observational study of a sample of interacting galaxies that result in poor starburst triggers}. \\
While, up to now, numerical simulations seem to support mainly the idea of a frequent occurence of tidally triggered central starbursts \citep{mih94,mih94b,mih96, spri00}, observational results draw a more complex scenario. Ultra Luminous Infrared Galaxies (ULIRGs) indicate centrally concentrated star formation \citetext{but see also \citealp{comb06} for an example of a moderate redshift ULIRG with a more extended CO emission}, while a certain number of interacting galaxies presents a more extended star formation. In the Antennae galaxies \citep{wang04}, for example, the most intense star formation regions are located between the two galaxies, and the late-type galaxy NGC 275 in the Arp 140 system, despite being close to a merger with its companion NGC 274, does not show evidence of an enhanced or centrally concentrated star formation, the brightest emission from star formation tracers coming from an off-center region \citep{cull06}.\\

Figs.\ref{nwsmergSbgSb09ret00}, \ref{gasmergSbgSb09ret00}, \ref{nwsmergSbgSb09dir00}, \ref{gasmergSbgSb09dir00}, \ref{nwsmergE0gSd03dir00} and \ref{gasmergE0gSd03dir00} show a comparison of most intense star formation regions and gas maps for three mergers of the sample. Far to represent the whole sample, they clearly suggest that the tidally triggered star formation process can manifest in a variety of  ways. The retrograde encounter/merger between two late type Sb galaxies leads to a strong enhancement in the total\footnote{i.e. evaluated on the total gas mass of the galaxy.} SFR (about 20 times that of the isolated case), the site of most intense star formation being strongly centrally concentrated. The same encounter, this time involving galaxies in direct orbits, leads to the star formation maps shown in Fig.\ref{nwsmergSbgSb09dir00}: in this case the central galaxy regions are sites of intense star formation in the final phases of the merging process, but, this time, the tidal tails and the bridge connecting the two galaxies contribute also to the production of new stars, as well as circumnuclear rings\footnote{Note that the presence of an ILR stops the gas inflow. This, together with the tidal gas removal, explains why SFRs tend to be lower in direct orbits,  as we will see later on.}. The situation drastically changes in  Fig. \ref{nwsmergE0gSd03dir00}, where the star formation maps resulting from an encounter between an early-type elliptical and a late-type Sd galaxy are shown. In this case, the first pericenter passage between the two galaxies is highly disruptive for the gSd, which develops two giant tails, that acquire a high fraction of gas mass, and that are also site of intense, local, and clumpy star formation. Note that the star forming clump in the upper left part of the panels in Fig.\ref{nwsmergE0gSd03dir00} corresponds to the center of the elliptical galaxy, that has acquired part of the gas mass of the companion during the interaction. \\ 
%%%the following  figures are discussed in the next Section
%-----------------------------------------------------------------
\begin{figure}[h]
  \centering
  \includegraphics[width=12.2cm,angle=270]{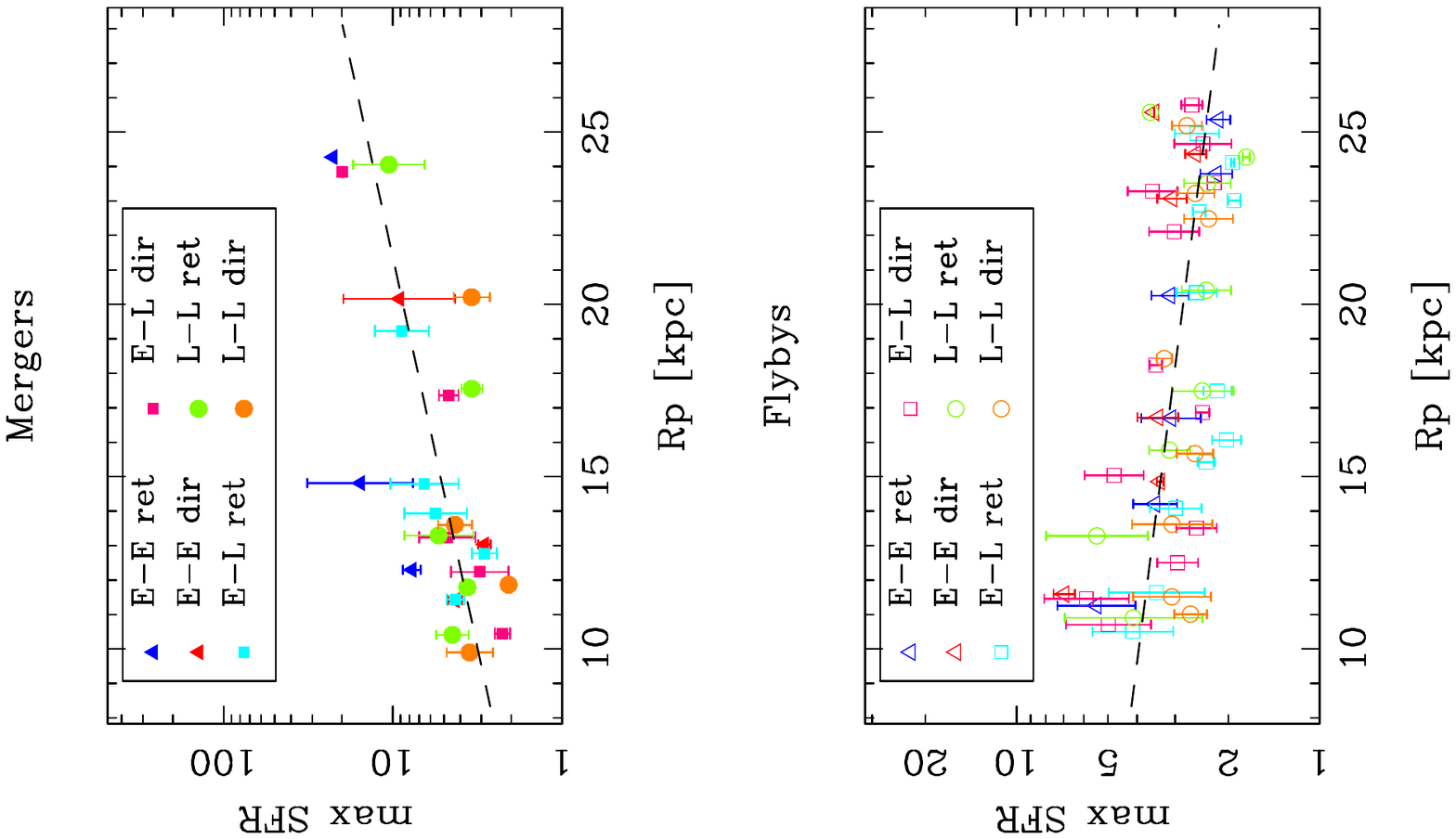}

  \caption{Maximum star formation rate, relative to the isolated case, as a function of the first pericenter distance between the two galaxies , for flybys (lower panel) and mergers (upper panel). Different symbols correspond to different morphologies of interacting galaxies, as explained in the legends. The dashed lines represent the best linear least-square fits. Error bars represent the standard error of the mean. }
  \label{psfrmax-rp}%
\end{figure}
On average, the gas mass being spread at great distances from the galaxy center, the gas density is obviously lower than the one obtained in the above cited encounter. This lead to a total SFR which is sensibly lower than the one obtained in the other two mergers: only 3 times higher than the isolated case.

\subsection{How does the maximum SFR depend on the total amount of gas available in the galaxies?}\label{does1}

Before moving on to describe in details the gas and stars dynamical evolution during the encounters, in order to try to understand the physical processes that lead to an enhancement in the SFR and in the ISFR, with respect to the isolated cases, we want to understand whether, and to what extent, the SFR peaks depend on the amount of gas available in the galaxy. In other words, are the different SFR peak values shown in Figs. \ref{sfrint}, \ref{sfrfly}, \ref{max-orb}, \ref{pisfr-orbgtutte} and \ref{psfe}  due to a different gas content in the galaxy at the moment of the burst? \\
To investigate this, we evaluated the gas content\footnote{The gas content considered is the total amount of gas,  i.e. the gas can be present both in the tails or in the main body of the galaxy. } (relative to that at the beginning of the simulation) just before the burst, i.e. for flybys we evaluated the amount of gas in the galaxy 50 Myr before the pericenter passage, while for mergers, the quantity is evaluated 50 Myr before the coalescence between the two galaxies. In Fig.\ref{psfrmax-gas} the maximum SFR is plotted, as a function of the gas content, expressed in units of the initial galactic gas mass, for all the encounters. 
Flybys and mergers locate in two different regions of the plots: 
for flybys, the pericenter passage occurs in the initial phases of evolution, when the gas content of the disk galaxies is similar to the initial ones, while the mergers are located in a region of space where the amount of gas mass is a fraction between 0.2 and 0.4 of the initial gas mass for elliptical-spiral encounters, while it is in a somewhat larger range (0.1-0.6) for spiral-spiral ones.\\ 
In both cases (flybys and mergers) it is evident that one can have large differences in the maximum SFR, even when the amount of fuel available is nearly the same: this is the case of flybys involving a gE0 and a gSb in direct orbits, for example, where one can have a factor of 4 in the maximum SFRs, even if the gas mass is comparable. Even more striking is the case of mergers where, with an available gas mass which is about  33$\%$ of the initial value, there are encounters which lead to a maximum SFR 60 times the isolated case, while others enhance the SFR only of a factor of 5.\\

In conclusion, Fig.\ref{psfrmax-gas} clearly shows  that \emph{the quantity of fuel still available in the galaxy is not the main parameter which influences the SFR in the  burst phase}.  

\subsection{How does the maximum SFR depend on $R_p$, $V_p$, $t_{enc}$ and on the effect of tidal forces?}\label{does2}

\begin{figure}
  \centering
  \includegraphics[width=12cm,angle=270]{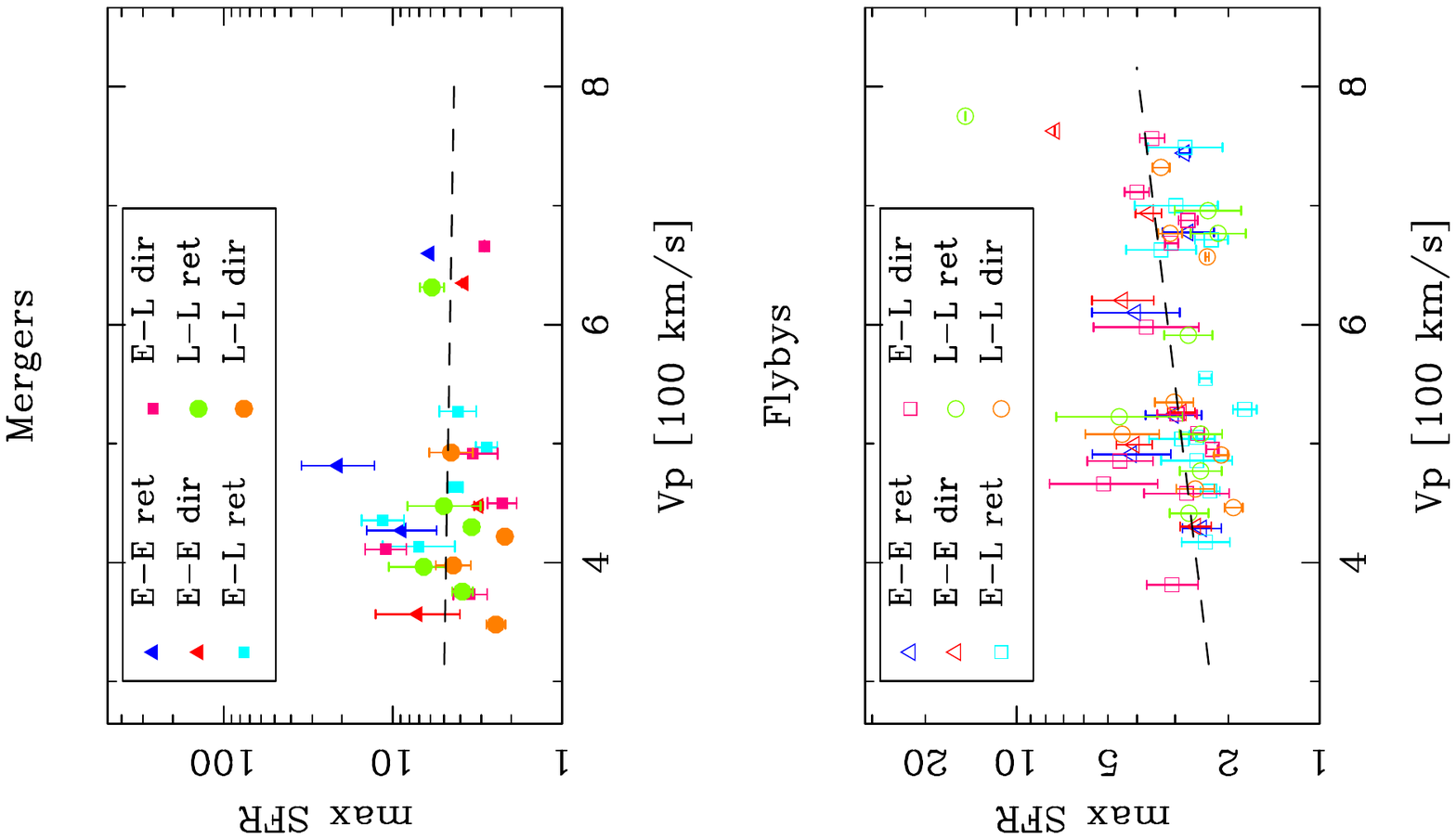}

  \caption{Maximum star formation rate, relative to the isolated case, as a function of the relative velocity at pericenter, for flybys (lower panel) and mergers (upper panel). Different symbols correspond to different morphologies of interacting galaxies, as explained in the legends. The dashed lines represent the best linear least-square fits. Error bars represent the standard error of the mean. }
  \label{psfrmax-vp}%
\end{figure}
%-----------------------------------------------------------------
\begin{figure}
  \centering
  \includegraphics[width=12cm,angle=270]{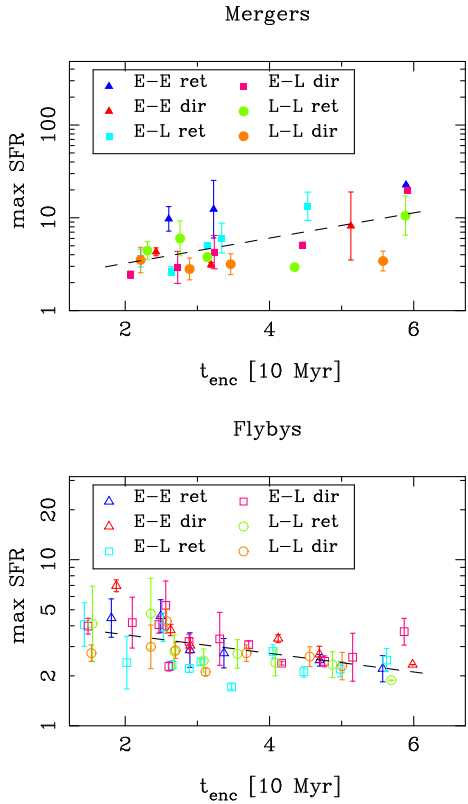}

  \caption{Maximum star formation rate, relative to the isolated case, as a function of the characteristic time of the encounter $t_{enc}$, for flybys (lower panel) and mergers (upper panel). Different symbols correspond to different morphologies of interacting galaxies, as explained in the legends. The dashed lines represent the best linear least-square fits. Error bars represent the standard error of the mean. }
  \label{psfrmax-tenc}%
\end{figure}
\begin{figure}
  \centering
  \includegraphics[width=12cm,angle=270]{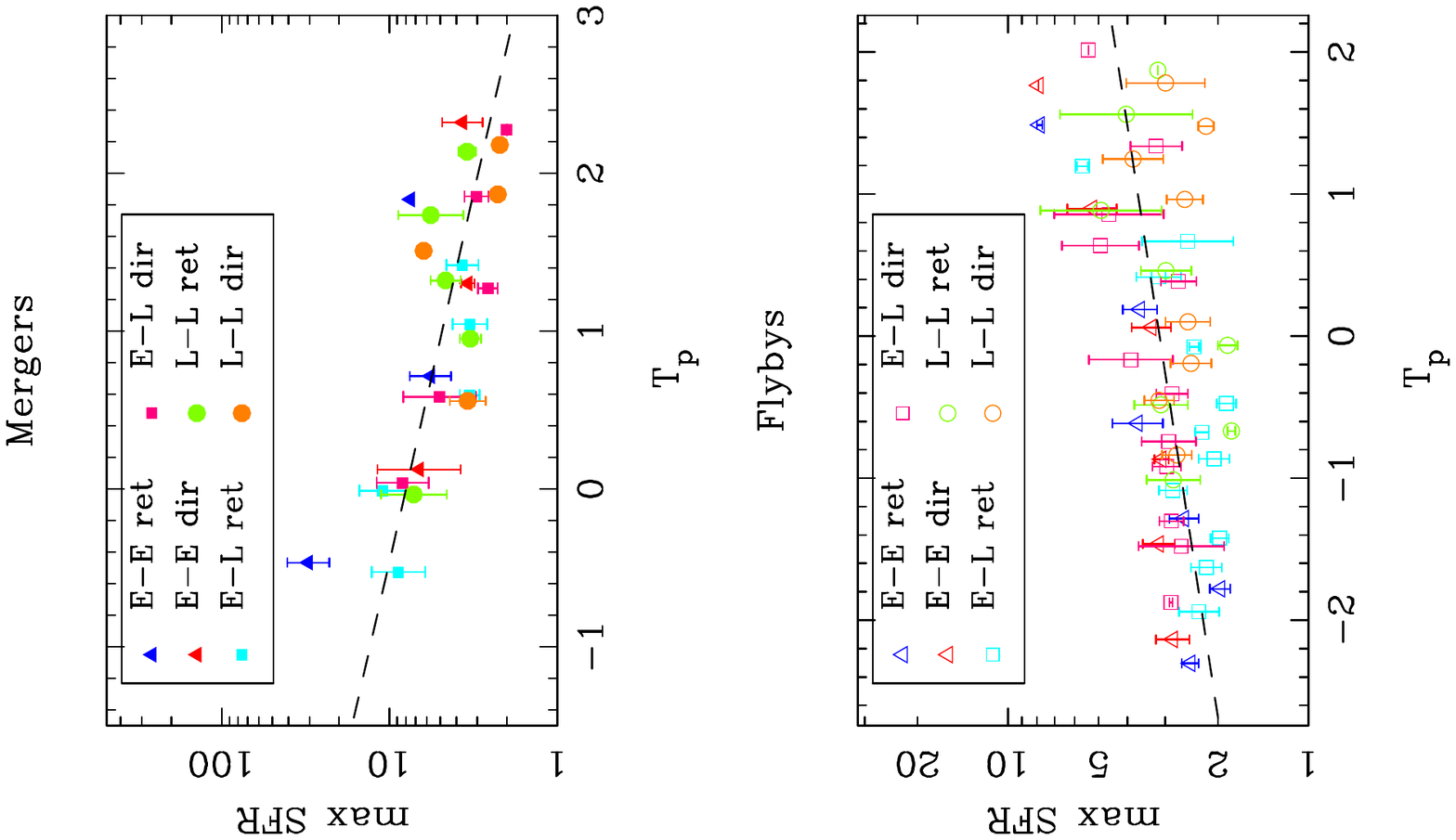}

  \caption{Maximum star formation rate, relative to the isolated case, as a function of the tidal parameter $T_{p}$ (see text), for flybys (lower panel) and mergers (upper panel). Different symbols correspond to different morphologies of interacting galaxies, as explained in the legends. The dashed lines represent the best linear least-square fits. Error bars represent the standard error of the mean. }
  \label{psfrmax-tp}%
\end{figure}
\begin{figure}
  \centering
  \includegraphics[width=12cm,angle=270]{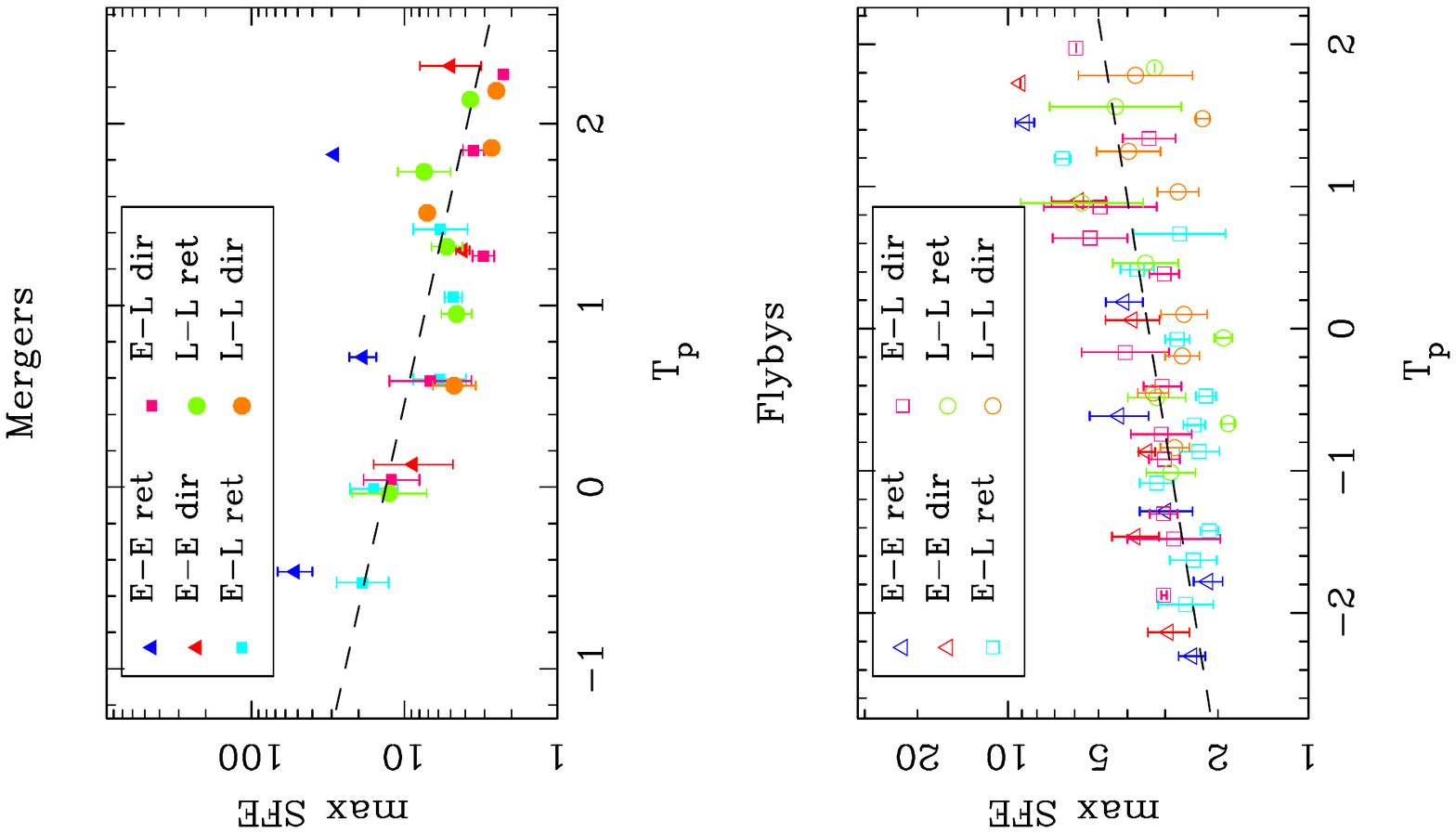}

  \caption{Maximum star formation efficiency, relative to the isolated case, as a function of the tidal parameter $T_{p}$ (see text), for flybys (lower panel) and mergers (upper panel). Different symbols correspond to different morphologies of interacting galaxies, as explained in the legends. The dashed lines represent the best linear least-square fits. Error bars represent the standard error of the mean. }
  \label{psfemax-tp}%
\end{figure}

In the previous Section, we have found that the maximum SFR in interacting and merging galaxies does not depend mainly on the total amount of gas mass available in the main body of the system and in the tails.
A first indication of the fact that the encounter geometry affects the enhancement of star formation has been found in Section \ref{sfrpairs}, comparing retrograde and direct encounters: we have seen that usually  retrograde interactions/mergers are more efficient than direct ones in converting gas to stars. 
 But it is still clear that other parameters must play a role in the evolution of the SFR in interacting galaxies: even considering only retrograde encounters between galaxies with the same morphology, in some cases the increase in the SFR relative to the isolated evolution is only modest, while other events lead to a SFR 60 times greater than that of the isolated case. \\
So, the aim of this Section is to explore the dependence of the maximum SFR  (occurring in the coalescence phase for mergers and soon after the pericenter passage for flybys) on other parameters, such as: 
\begin{itemize}
\item the distance at the first pericenter passage $R_p$;
\item the relative velocity $V_p$ at $R_p$\footnote{$R_p$ and $V_p$ have been evaluated using the definitions of galaxy centers and velocities given at the beginning of Sect.\ref{results}.};
\item the characteristic time of the encounter, $t_{enc}=\frac{R_p}{V_p}$
\item the tidal parameter $T_p$ of the galaxy pair, defined as\footnote{Note that in the following definition of $T_p$ no dependence on the galaxy mass ratio is proposed. The sample being composed only by equal mass galaxies, this has no influence on the present study. A possible formulation for unequal mass encounters could consist into introducing a weight $g=M_i/(M_i+M_{comp})$ in the expression for $T_{p,\ i}$ , i.e. $T_{p,\ i}=log_{10}[g(M_{comp}/M_{i})\times(D_{i}/R_p)^3]$.}
\begin{equation}\label{tptot}
T_{p}=T_{p,1}+T_{p,2},
\end{equation}
where
\begin{equation}\label{tp}
T_{p,i}=log_{10}\left[\frac{M_{comp}}{M_{i}}\left(\frac{D_{i}}{R_p}\right)^3\right], \ \ i=1,2
\end{equation}
 quantifies the effect of tidal forces at pericenter passage $R_p$, suffered by a galaxy of mass $M_{i}$ and scalelength $D_{i}$, due to the interactions with a companion galaxy whose mass is $M_{comp}$ \citep[see also][]{bour05}. As a measure of the galaxy scalelength $D_{i}$ we used the radius containing $75\%$ of the total (baryonic + dark matter) mass of the system;
\item the parameter $T_p/t_{enc}$ that gives an estimate of the effect of tidal forces exterted on the pair, per unit of time. 
\end{itemize}

We want to recall that, while for flybys this analysis relates the SFR at pericenter passage with quantities evaluated at the pericenter too, this is not the case of mergers, looking in this case for possible existing correlations between the SFR in the coalescence phase and physical quantities evaluated at the first pericenter passage. 

\subsubsection{Maximum SFR versus $R_p$}

\begin{figure}
  \centering
  \includegraphics[width=12cm,angle=270]{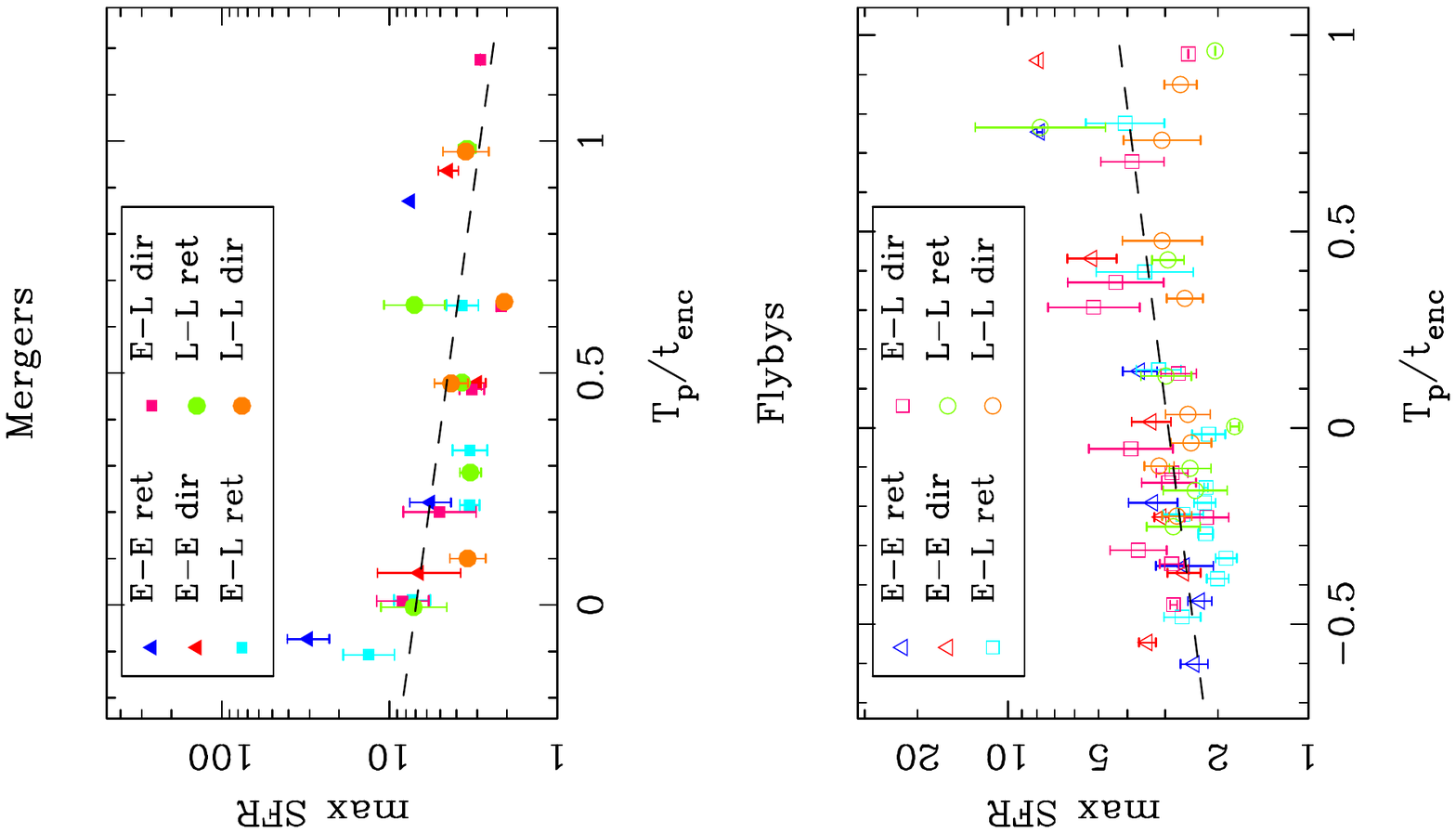}

  \caption{Maximum star formation rate, relative to the isolated case, as a function of $T_{p}/t_{enc}$ (see text), for flybys (lower panel) and mergers (upper panel). Different symbols correspond to different morphologies of interacting galaxies, as explained in the legends. The dashed lines represent the best linear least-square fits. Error bars represent the standard error of the mean. }
  \label{psfrmax-tpsutenc}%
\end{figure}

Fig. \ref{psfrmax-rp} shows the maximum SFR , as a function of the distance $R_p$ at first pericenter passage, for flybys (lower panel) and mergers (upper one).\\
It is not surprising to see that for flybys there is a tendency to have increasing star formation levels as the pericenter separation of galaxies in the pair diminishes. More interesting is the case of mergers, where a positive correlation (+0.7) between these two quantities is found, so that galaxies that at first passage are more distant are also those that suffer the most intense burst of star formation in the merging phase. 

\subsubsection{Maximum SFR versus $V_p$}

No correlation is found  between SFR maximum and the relative velocity of the two galaxies at the first pericenter passage  for the merger sample, while a weak correlation (+0.34) exists for flybys (see Fig.\ref{psfrmax-vp}).

\subsubsection{Maximum SFR versus $t_{enc}$}

As it is shown in Fig.\ref{psfrmax-tenc}, the duration of the encounter, so, ultimately, the duration of the perturbation induced by the companion galaxy, is a parameter that influences the peak of star formation (the absolute value of the correlation coefficient is about  0.6 for both flybys and mergers). And, as previously found for $R_p$, the two samples of flybys and merger galaxies show a completely different behaviour: indeed, while the maximum SFR increases as $t_{enc}$ diminishes for flybys, the opposite is the case for mergers, having a tendence to have higher bursts of star formation at increasing $t_{enc}$.
This is exactly the same trend shown in Fig. \ref{psfrmax-rp}, because of the fact that flybys with small  pericentric distances must also have the  highest relative velocities, and so, ultimately, small $t_{enc}$.

\subsubsection{Maximum SFR versus $T_{p}$}\label{sfrtp}

For mergers, a strong anti-correlation (-0.72) is found when plotting the maximum SFR versus the tidal parameter $T_p$, defined as the sum of the tidal parameters $T_{p}$ exerted by the two galaxies. 
The clear trend in Fig. \ref{psfrmax-tp} is that \emph{galaxy pairs that suffer intense tidal effects at first pericenter passage tend to have lower star formation rates in the merging phase}. The same behaviour is found when comparing the SFE at the merger epoch with $T_p$ (see Fig. \ref{psfemax-tp}). The opposite trend stands for flybys, i.e. \emph{at pericenter, galaxy pairs that suffers intense tidal effects tend to have also higher star formation rates}.

\subsubsection{Maximum SFR versus $T_{p}/t_{enc}$}

The same trend is found  when comparing the SFR at the merger with $T_p/t_{enc}$, which gives an estimate of the perturbation due to tidal forces per unit time.
With a correlation coefficient $r=-0.55$, it shows that the maximum SFR is somewhat affected by the tidal forces exerted by the companion galaxy per unit time. We want to recall that our star formation law is only density dependent, so that an increase in the SFR corresponds to an increase in the local gas density. In this sense, the anticorrelation found between the amplitude of the burst and the tidal forces exerted per unit of time can be understood if one considers that the strongest the forces, the greatest the quantity of galaxy mass (and so, ultimately, also of gas material) removed from the system, as effect of energy injection (see next Section for a deeper discussion).

\subsection{Why does the SFR in the merging phase depend on $T_p$?}\label{tidal}

\begin{figure}
  \centering
  \includegraphics[width=4.5cm,angle=270]{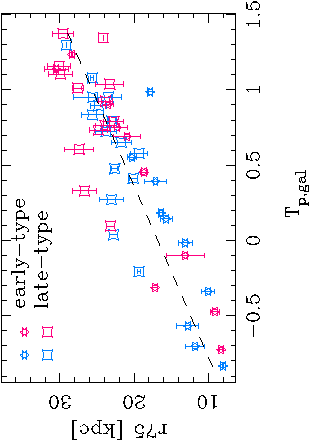}
  \caption{Radius containing the 75\% of the total (baryons+dark matter) mass of the galaxy versus the galaxy tidal parameter $T_{p,gal}$. Both quantities have been evaluated just after the first pericenter passage. Different symbols correspond to early and late-type galaxies, as explained in the Figure. Red color refers to direct encounters, blue to retrograde ones. The analysis has been restricted to galaxies in the merger sample (see text). The dashed lines represent the best linear least-square fits. Error bars represent the standard error of the mean. }
  \label{pr75-tp}%
\end{figure}

The goal of this Section is to further investigate the relation found in Fig. \ref{psfrmax-tp} between the value of the SFR peak in the merging phase and effects of tidal forces on the pair, evaluated at the first pericenter passage.
In all the previous sections, we have always refered our analysis to the couple of interacting galaxies: for example, the SFR is the total SFR, to which both galaxies of the pair contribute; the tidal parameter $T_p$ has been evaluated following Eq. \ref{tp}, so summing the tidal effects suffered by the two galaxies.
In this  Section, we will ``decompose'' our analysis, evaluating some physical quantities directly on each galaxy of the pair. This in order to have a more detailed picture of the role played by each component in the evolution of the interacting system.\\
Being interested in explaining the relation found for mergers, here we restrict our analysis to the sample of merger galaxies, i.e. to all the galaxies involved in interactions that lead to mergers in 3 Gyr of evolution.\\
First of all, for each galaxy in this sample, we evaluate the radius $r_{75}$ containing $75\%$ of the total mass (gas+stars+dark matter) of the system, just after the pericenter passage with the companion, this in order to quantify the expansion experienced by the galaxy, as an effect of the encounter. In Fig.\ref{pr75-tp}, $r_{75}$ is plotted versus the tidal parameter $T_{p, gal}$, which now is evaluated on the single galaxy and not on the pair. For simplicity, we have grouped galaxies into two subsamples: early-type systems, including gE0 and gSa, and late-type ones, including gSb and gSd galaxies.\\ 

\begin{figure}
  \centering
  \includegraphics[width=4.5cm,angle=270]{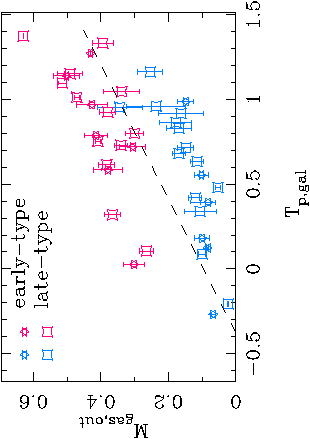}
  \caption{Gas mass outside 20 kpc from the galaxy center, versus the galaxy tidal parameter $T_{p, gal}$. $M_{gas,out}$ has been evaluated just after the pericenter passage and it has been normalized to the gas mass present, at the same time,  in the disk of the corresponding galaxy evolving alone. Different symbols in the figure correspond to early and late-type galaxies, as explained. Red color refers to direct encounters, blue to retrograde ones. The analysis has been restricted to galaxies in the merger sample (see text). The dashed lines represent the best linear least-square fits. Error bars represent the standard error of the mean. } 
  \label{pgasout}%
\end{figure}

Not surprisingly, the figure shows a strong correlation between these two quantities, indicating that \emph{the stronger the interaction at the pericenter passage, the greater the subsequent expansion of the outer parts of the system}. The radius $r_{75}$ is computed including all the galaxy components, but it is likely that a similar relation holds  for the gaseous component only, in the sense that the gas mass ejected into tidal tails grows as tidal effects on the galaxy become stronger. This is shown in Fig.\ref{pgasout}, where  $M_{gas,out}$, the gas mass outside 20 kpc from the galaxy center, is plotted versus $T_{p, gal}$.  $M_{gas,out}$ has been evaluated just after the pericenter passage and it has been normalized to the gas mass present, at the same time,  in the disk of the corresponding galaxy evolving alone.  \\

To summarize, after the pericenter passage, the amount of gas in the galactic disk is depleted by two complementary effects:
\begin{enumerate}
\item an increase in the star formation rate in the disk, which accelerates the gas consumption;
\item the ejection of gas material from the disk into tidal tails.
\end{enumerate}

\begin{figure}
  \centering
  \includegraphics[width=4.5cm,angle=270]{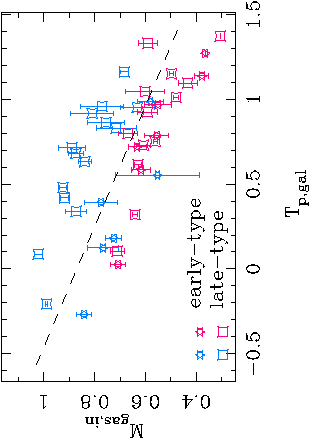}
  \caption{Gas mass inside 20 kpc from the galaxy center, versus the galaxy tidal parameter $T_{p, gal}$. $M_{gas,in}$ has been evaluated just after the pericenter passage and it has been normalized to the gas mass present, at the same time,  in the disk of the corresponding galaxy evolving alone. Different symbols in the figure correspond to early and late-type galaxies, as explained. Red color refers to direct encounters, blue to retrograde ones. %The analysis has been restricted to galaxies in the merger sample (see text). 
The dashed lines represent the best linear least-square fits. Error bars represent the standard error of the mean. }
  \label{pgasin}%
\end{figure}

\begin{figure}
  \centering
  \includegraphics[width=4.5cm,angle=270]{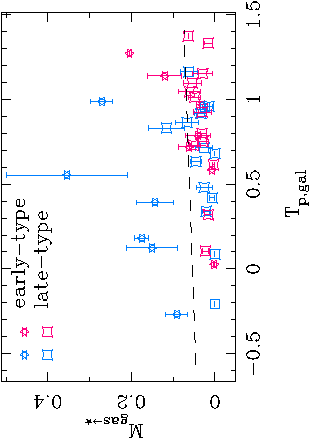}
  \caption{Plot of $M_{gas\to \star}$ versus the galaxy tidal parameter $T_{p, gal}$. $M_{gas\to \star}$ represents the integrated gas mass transformed into star from the time $t=0$ to the time $t=t(R_p)$. As in Figs.\ref{pgasout} and \ref{pgasin}, this mass has been  normalized to the gas mass present, at the same time,  in the disk of the corresponding galaxy evolving alone. Different symbols in the figure correspond to early and late-type galaxies, as explained. Red color refers to direct encounters, blue to retrograde ones. Error bars represent the standard error of the mean.  The analysis has been restricted to galaxies in the merger sample (see text).}
  \label{pgastar}%
\end{figure}

The role these two phenomena play in depleting the gas mass in the disk is shown in Figs. \ref{pgasin} and \ref{pgastar}. \\
The first figure shows  $M_{gas,in}$, the gas mass  inside 20 kpc from the galaxy center, versus $T_{p, gal}$. As in Fig. \ref{pgasout}, $M_{gas,in}$ has been evaluated just after the pericenter passage and it has been normalized to the gas mass present, at the same time,  in the disk of the corresponding galaxy evolving alone.\\
 The second figure shows the dependence on  $T_{p,gal}$ of $M_{gas\to \star}$, the integrated gas mass trasformed into star from the time $t=0$ to the time $t=t(R_p)$\footnote{It is the time corresponding to the pericenter passage.}. This quantity  also has been normalized to the gas mass present, at the same time,  in the disk of the corresponding galaxy evolving alone.\\
Comparing these two Figures with Fig.\ref{pgasout}, it is evident that usually the disk gas mass is decreased mainly because of the ejection of a conspicuous gas fraction far out from the disk, while the encounter-enhanced star formation contributes only modestly to clear it out.
Obviously, part of the gas mass lost in the tails, or acquired by the companion galaxy, can be reaccreted in the last stages of the merging event, thus contributing to the fuel available in the burst phase. But this is not the case for all the gas mass lost from the disk. 
In this sense, Fig.\ref{pgasout} contributes to explain the trend found in Fig.\ref{psfrmax-tp} (top panel).

\subsection{Toward a formulation for the SFE}\label{toward}

For practical use, it is convenient to have a fitting formula for the estimation of the SFE, even if limited to planar encounters.
The aim of this Section is to propose a possible formulation, for mergers and flybys, as a function of: 1) the type of the encounter (E-E, E-L, L-L); 2) the orbital spin, if parallel or antiparallel to the galaxies angular momentum (i.e. direct or retrograde orbit); 3) the pericenter distance $\mathrm{R_{2b}}$ of the Keplerian orbit of two points with masses equal to the masses of the interacting systems.

 In other terms, we are looking for a function $f(\mathrm{morph, spin, R_{2b}})$ so that 
\begin {equation}\label{funct}
log_{10}\mathrm{SFE}_{max}=f(\mathrm{morph, spin,R_{2b}}).
\end{equation}
As in the previous Sections, for mergers and flybys, the maximum SFE is relative to that of the corresponding galaxies evolved in isolation.\\

In Section \ref{sfrtp} we have shown the strong correlation existing between the maximum SFE and the tidal parameter $\mathrm{T_p}$ at the pericenter, giving a global fitting formula, which takes into account all the various morphological galaxy types.
In general, this strong correlation between $\mathrm{T_p}$  and SFE still holds  when grouping galaxies in different classes depending on their morphological type and the orbital spin, direct or retrograde. 
This suggests to adopt a right-hand side in Eq.\ref{funct} containing a parameter that could approximate $T_p$. Note indeed that, for a practical use, we are looking for a formulation for SFE that could be expressed only in terms of the initial parameters of the simulated encounter and that does not depend on quantities that require  an \emph{ad hoc} simulation to be evaluated. \\
We remind the reader that the tidal parameter $T_p$ at pericenter has been defined in Eq.\ref{tptot} as the sum of the tidal parameters $T_{p,1}$ and $T_{p,2}$ relative to the two galaxies involved in the interaction, i.e.
\begin{eqnarray}
T_{p}&=&log_{10}\left[\frac{M_{2}}{M_{1}}\left(\frac{D_{1}}{R_p}\right)^3\right]+log_{10}\left[\frac{M_{1}}{M_{2}}\left(\frac{D_{2}}{R_p}\right)^3\right]=\\
&&log_{10}\left[\left(\frac{D_{1}D_{2}}{{R_p}^2}\right)^3\right],
\end{eqnarray}
 $R_p$ being the pericenter distance and $D_1$ and $D_2$ the galaxies scalelength.  $R_p$ was evaluated as the minimum distance between the two galaxy density centers and  both $D_1$ and $D_2$ were approximated with the radius containing  $75\%$ of the total galaxy mass and where evaluated at the pericenter passage.\\
On the contrary, the suggested approximation for $T_p$ will make use of all quantities yet available \emph{ab initio}, i.e.

\begin{equation}\label{x}
T_p \approx x=log_{10}\left[\left(\frac{d_{1}d_{2}}{{R_{2b}}^2}\right)^3\right],
\end{equation} 
 $R_{2b}$ being the pericenter distance of the unperturbed Keplerian orbit of two bodies, having masses equal to the two galaxy masses, and $d_{1}$ and $d_{2}$ being the radii containing $75\%$ of the total mass of the unperturbed galaxies.\\

The suggested formulation for the SFE is the following\footnote{The  sample being composed only by giant galaxies, in the following formula no dependence on the galaxy masses is proposed. We refer the reader to following papers for a more general formulation, that could take into account also the mass ratio of the interacting galaxies.  }:
\begin{equation}\label{formula}
log_{10}\mathrm{SFE}_{max}=A\ log_{10}\left[\left(\frac{d_{1}d_{2}}{{R_{2b}}^2}\right)^3\right]+B,
\end{equation}
being $A=A(\mathrm{morph, spin})$ and $B=B(\mathrm{morph, spin})$ constants depending on the morphology of the two galaxies involved in the interaction and on the orbital spin. Note that also the $x$ term depends on galaxy morphologies, via $d_{1}$ and $d_{2}$.\\
%%%%%%%%the following table is discussed in the next section
% table
   \begin{table}
      \caption[]{Values for the A and B constants in Eq.\ref{formula}, for mergers and flybys.}
         \label{AB}
	 \centering
         \begin{tabular}{ccccc}
            \hline\hline
	    & \multicolumn{2}{c} {Mergers}& \multicolumn{2}{c} {Flybys}\\
	    \hline
	    & A & B & A & B \\
            \hline
	    E-E ret & -0.098 & 1.557 & 0.135 & 0.653\\
	    E-E dir & -0.126 & 0.826 & 0.115 & 0.724\\
	    E-L ret & -0.202 & 1.089 & 0.078 & 0.482\\
	    E-L dir & -0.098 & 0.779 & 0.024 & 0.543\\
	    L-L ret & -0.121 & 0.954 & 0.027 & 0.461\\
	    L-L dir & -0.026 & 0.621 & 0.004 & 0.467\\
            \hline
         \end{tabular}
   \end{table}

\begin{figure}[h]
  \centering
  \includegraphics[width=6.3cm,angle=270]{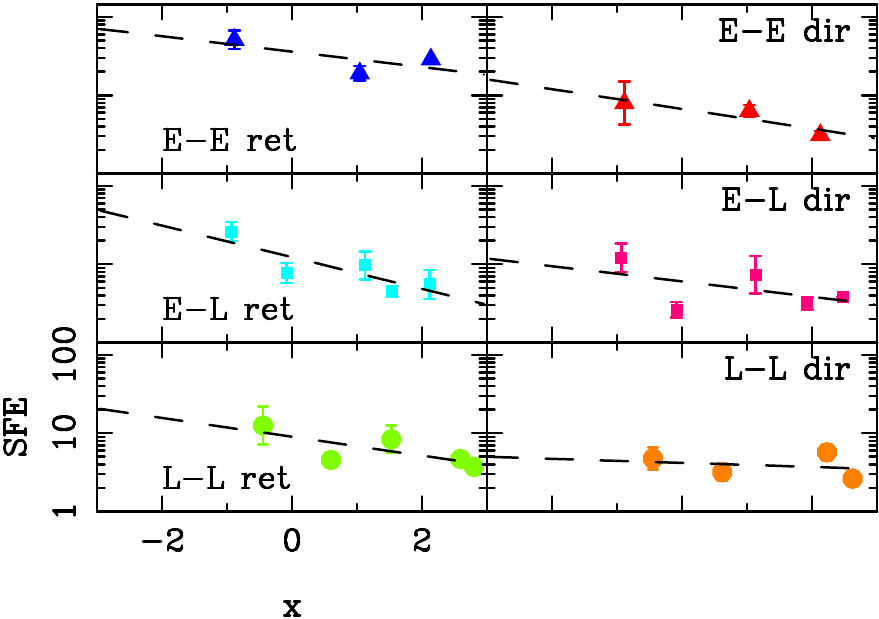}
  \caption{SFE versus the parameter $x$ defined in Eq.\ref{x} for mergers. Different panels refer to different classes for the encounters, depending on the morphology of the interacting galaxies and on the orbital spin. Error bars represent the standard error of the mean.  The dashed line in each panel represents the proposed formulation for the SFE (see Eq.\ref{formula}).}
  \label{plawmer}%
\end{figure}

\begin{figure}[h]
  \centering
  \includegraphics[width=6.3cm,angle=270]{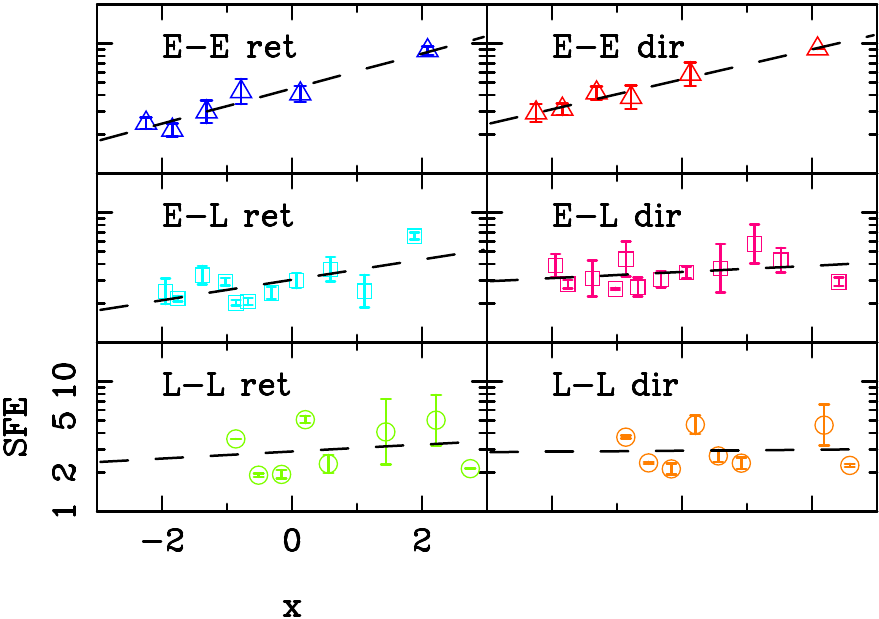}
  \caption{SFE versus the parameter $x$ defined in Eq.\ref{x} for flybys. Different panels refer to different classes for the encounters, depending on the morphology of the interacting galaxies and on the orbital spin. Error bars represent the standard error of the mean.  The dashed line in each panel represents the proposed formulation for the SFE, described (see Eq.\ref{formula}).}
  \label{plawfly}%
\end{figure}

In Table \ref{AB}, $A$ and $B$ are given, both for mergers and flybys, and for different classes (E-E, E-L, L-L, in retrograde or direct orbits). Their values have been obtained performing a linear least-square fit between the $x$ values and the corresponding values of $log_{10}\mathrm{SFE}_{max}$ resulting from simulations.\\
A comparison between the curves
\begin{equation}\label{theo}
y=A\ log_{10}\left[\left(\frac{d_{1}d_{2}}{{R_{2b}}^2}\right)^3\right]+B
\end{equation}
and the 'experimental' data for the SFE, for mergers and flybys, is shown in Figs.\ref{plawmer} and \ref{plawfly}, respectively. The proposed formulation for the SFE reproduces the data quite well, with the modulus of the relative errors quite uniformly distributed with a mean value that is always below $30\%$, with the only exception of Early-Late type direct mergers and Late-Late type retrograde flybys, where it is $\approx 50\%$.


\subsection{Evolution in the $(\Sigma_{gas}, \Sigma_{SFR})$ plane}\label{globschmidt}

%%%the following figure is discussed in the next Section
%---------------------------------
\begin{figure*}
  \centering
  \includegraphics[width=13cm,angle=270]{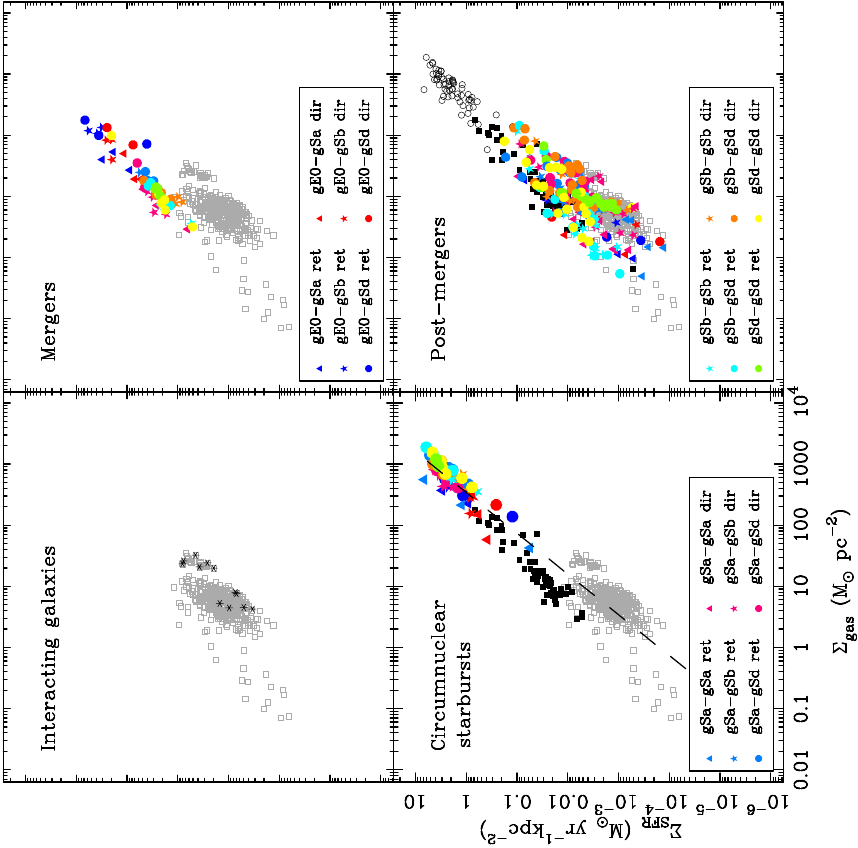}
  \caption{The global Schmidt law in galaxies. Upper left panel:  $(\Sigma_{gas}, \Sigma_{SFR})$ relation for interacting galaxies (grey empty squares).  For comparison, also isolated galaxies are shown (stars). Upper right panel:  $(\Sigma_{gas}, \Sigma_{SFR})$ relation for merging galaxies. Different symbols in this and in the next two panels are explained in the legends. For comparison, the interacting galaxy sample is shown with grey empty squares.  Lower left panel: circumuclear starburst.  For comparison, interacting galaxies (grey empty squares) and mergers (black squares) are also shown.  The dashed line represent a ``global'' Schmidt star formation law of the form $\Sigma_{SFR}\propto{\Sigma_{gas}}^{1.4}$ \citep{ken98}, and it is shown for a comparison with the data. Lower righ panel: $(\Sigma_{gas}, \Sigma_{SFR})$ relation for post-merger galaxies.  For comparison, interacting galaxies (grey empty squares), mergers (black squares) and circumnuclear starburst (black empty circles) are also shown.
}
  \label{pschmidt_tot}%
\end{figure*}
%---------------------------------

In the previous sections, we have seen that the star formation rate in the merging phase anticorrelates with the intensity of the tidal interaction at first pericenter passage and we explained this in terms of gas removal from the galactic disk of the two galaxies, just after the encounter.
In this picture, the next step is to verify if the SFR during interactions and in the coalescence phase correlates with the quantity of gas material available in the galactic disk.\\

Several observations of normal galaxies suggest that the SFR on global scales is well represented by a Schmidt law of the form $\Sigma_{SFR}=A{\Sigma_{gas}}^N$, $\Sigma_{gas}$ being the gas density weighted inside the galactic optical radius and $\Sigma_{SFR}$ the disk-averaged SFR surface density. The best fitting slope $N$, obtained with a conventional least squares fit, is about 1.4 \citep[see][]{ken98}. Even if with large variation for a given value of $ \Sigma_{gas}$, these data show that a correlation between SFR and gas density exists, which is also largely independent of galaxy type. A similar correlation, with a comparable $N$ slope, exists  for starburst galaxies, if averaging the SFR density and the gas density on the inner (1-2 kpc) galactic region, both the $ \Sigma_{gas}$ and the $\Sigma_{SFR}$ values  being in this case 1-4 orders of magnitude higher than those obtained for normal galaxies, thus suggesting that a large amount of gas is present in the central region to sustain the burst activity.\\
Other authors found some variations in the value of the $N$ slope: \citet{wb02} studing a sample of seven molecule rich spiral galaxies found $N\sim1.1-1.7$, depending on the correction for extinction in $H\alpha$ emission, in deriving the SFR; \citet{bois03} found $N\sim2.0$ for a sample of sixteen spiral galaxies; finally \citet{gs04} suggested a star formation law with a power-law index of 1.0 in terms of dense molecular gas content, studying a sample of 65 normal spirals, luminous infrared and ultraluminous infrared galaxies.\\

In this framework, it is interesting to study if our sample obeys a similar global relation, provided that on local scales our star formation recipe is a Schmidt type. In particular, we are interested in understanding if interacting and starburst galaxies follow a ``global'' Schmidt law of the type $\Sigma_{SFR}=A{\Sigma_{gas}}^N$, as found observationally for normal spiral galaxies \citep{ken98} and if, more generally, it is possible to trace a global evolution on the $\Sigma_{gas}-\Sigma_{SFR}$ plane, for pre-interaction, interacting, starburst galaxies and mergers.\\
To compute $\Sigma_{gas}$ and $\Sigma_{SFR}$, we evaluated the SFR and the gas amount inside a radius containing $85\%$ of the total visible (gas+stars) galaxy mass\footnote{Note that for interacting and starburst galaxies, due to the fact that part of the mass populates tidal tails, thus spreading for several hundreds kpc outside the galactic center, we restricted our analysis to  $85\%$ of the total visible (gas+stars) mass located inside 20 kpc from the galaxy center.} and then we calculated the disk-density relative quantities. For interacting galaxies, we evaluated the two quantities for each galaxy of the pair, while for starburst, circumnuclear starbursts and post-merger  $ \Sigma_{gas}$ and the $\Sigma_{SFR}$ are evaluated on the resulting merger.\\

The main findings of our analysis are summarized in Fig.\ref{pschmidt_tot}, which shows different galaxy samples (interacting, starbursts, circumnuclear starbursts and mergers) in the plane  $(\Sigma_{gas}, \Sigma_{SFR})$.\\
In this plane the interacting galaxies (upper left panel) lie in a region which extends over  2 orders of magnitude both in $\Sigma_{gas}$, from 0.1 to 30 $M_{\odot}pc^{-2}$, and in $\Sigma_{SFR}$ ($10^{-4}-0.02 M_{\odot}yr^{-1}kpc^{-2}$). No evident different behaviour is found for different morphological type, i.e. for early and late-type systems. This sample includes galaxies at very different stages of interactions: there are systems that are well before the first close passage, hundreds of kpc distant from the companion (in this sense they can be considered as isolated) and systems close to the merging phase. We eliminated from this sample (as in the isolated sample shown in the same panel for comparison) the transient initial burst phase, corresponding to the emergence of density waves in the galactic disks (see Figs. \ref{sfriso} and \ref{tot}). Note that the best linear fit reported in Fig.\ref{pschmidt_tot} for interacting galaxies would be steeper if the low  $\Sigma_{gas}- \Sigma_{SFR}$ values in this sample were not considered.\\

The starburst phase is represented in the upper right panel. The high correlation found for this sample (0.9) is remarkable. One can note that in this phase galaxies occupy an extended region in  $\Sigma_{gas}$, and reach higher levels of SFR density, with respect to interacting pairs. 
 Note also that  mergers resulting from encounters between an elliptical and a spiral have, in general, higher gas surface densities and higher star formation rate density with respect to spiral-spiral mergers.
 The situation changes if one restricts the analysis of starburst galaxies to circumnuclear regions, in the inner kpc region of the galaxy. In this case (lower left panel) the nature of the starburst event clearly emerges: the gas circumnuclear density is, on average, at least 100 times higher than that of the overall disk, thus giving rise to a strongly enhanced SFR, whose circumnuclear density is far greater than those of interacting and starburst galaxies as a whole. In this case, late-type spirals encounters show higher SFR densities, if compared to mergers involving an early-type system.\\

Once the starburst phase is over, the mergers evolve toward the lower left part of the  $(\Sigma_{gas}, \Sigma_{SFR})$ plane (as shown in the lower right panel in Fig.\ref{pschmidt_tot}), having at this point consumed most of their gas reservoirs.\\
This overall trend in the location of normal and starburst galaxies in the $(\Sigma_{gas}, \Sigma_{SFR})$ plane is confirmed also observationally \citep[see][Fig.9 left]{ken98}. Note that in this case highest values (for normal and merging galaxies) are reached both in the x and y axis, but it is just due to the fact that our simulations start with initial conditions typical of present day galaxies and then evolve for 3 Gyr in time, so that the amount of gas in the simulated systems is only initially comparable to that of galaxies in the local Universe. In particular, the position of isolated and interacting galaxies in Fig.\ref{pschmidt_tot} top left is a reflect of gas amount.
\\

The present study on the evolution of isolated, interacting and merging galaxies in the $(\Sigma_{gas}, \Sigma_{SFR})$ plane extends and enriches prior numerical works concerning star formation in isolated galaxies and in major interactions. \citet{spri00}, for example, studied the composite Kennicutt law for a small set of simulated isolated and merging galaxies, showing that his sample remarkably reproduces the global Schmidt law over a large dynamic range. But the accordance of the data with the global Schmidt law is a natural consequence of the chosen star formation and feedback recipes he adopted, being the free parameters chosen in order to satisfy Kennicutt's findings. \\
In an extensive study conducted to analyze how star formation depends on the adopted parametrization for feedback, \citet{cox06} showed that isolated galaxies satisfy the empirical Kennicutt law, when including a gas density threshold for star formation. Indeed, when averaging star formation rate and gas surface density within azimuthal annulli, and adopting not too high feedback parameters, they found a good accordance with the results in Fig.3 in \citet{ken98b}. Then, studying the Kennicutt law for the merger sample, using an azimuthal aperture of radius 2 kpc, they confirmed that also the interacting sample closely tracks the empirical star formation law. \\
But both these works %(that differ from our in some aspects, as in the adoption of a star formation threshold, and in the feedback modelization, designed to stabilize the gas disc and prevent excessive star formation due to disc instabilities) 
use a star formation and a feedback modelization designed to reproduce Kennicutt empirical law. In this sense, the present study investigates the  $\Sigma_{gas}-\Sigma_{SFR}$ correlation, in a more general context. Indeed, starting from a star formation recipe based on the local gas volume density (Eq.\ref{loc}), it is not immediately clear that a global (on 10-20 kpc scales) relation, based on gas surface density, must hold. Evidently, we do not claim to reproduce the slope found for the empirical law, but, neverthless, it is striking to see how a global (i.e. on kpc scales) $\Sigma_{gas}-\Sigma_{SFR}$ relation is satisfied, for isolated, interacting, merging and post-merger galaxies.     
 \begin{figure*}
  \centering
  \includegraphics[width=14cm,angle=270]{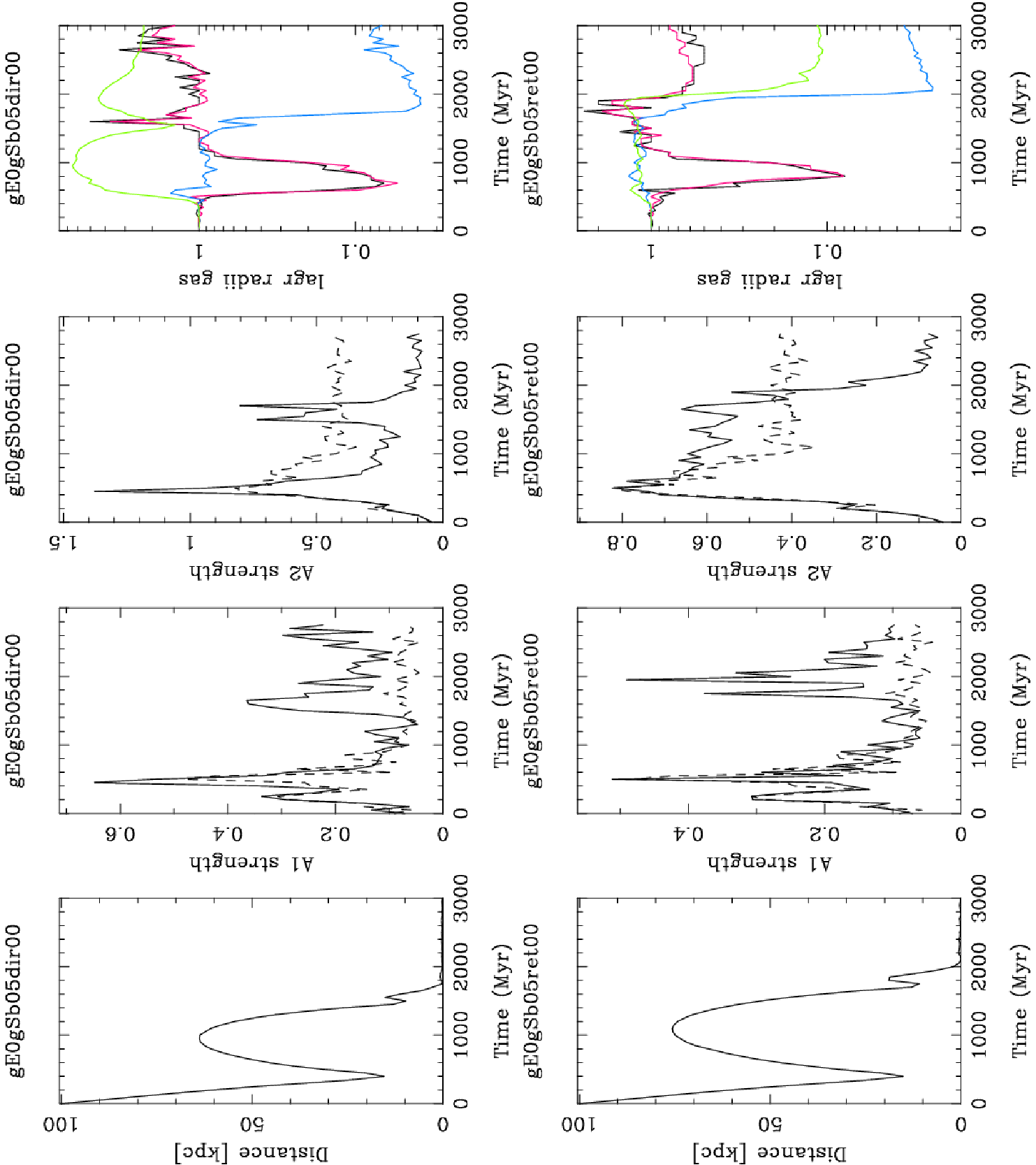}
  \caption{Gas dynamics during the interaction and subsequent merger between an elliptical and a spiral Sbc galaxy. The first row refers to the two galaxies in direct orbit, the second row to galaxies in retrograde interaction. First column: relative distance between the two galaxy centers versus time. Second column: Evolution of the $A_1$ strength, versus time (solid line). For comparison, the evolution of the $A_1$ asymmetry of the same galaxy evolving alone is shown (dotted line). Third column: Evolution of the $A_2$ strength, versus time (solid line). For comparison, the evolution of the $A_2$ asymmetry of the same galaxy evolving alone is shown (dotted line). Fourth column: Evolution with time of gas lagrangian radii containing $10\%$ (black), $25\%$ (red), $50\%$ (blue) and $75\%$ (green) of the gas mass are shown, in units of the corresponding gas radii of the isolated gSb galaxy.}
  \label{inflow}%
\end{figure*}

\subsection{Gas inflow}\label{gasinflow}

It is well established  that
the interaction with a companion usually destabilise the disk of a galaxy, which develops non axisymmetric structures (as spiral arms, bars), responsible of exerting torques on the gas material (Noguchi 1988, Barnes \& Hernquist 1996, Mihos \& Hernquist 1996, Combes 2001). In this way, the gas loses angular momentum and falls into the inner galactic regions, where a starburst takes place \citep[see ][for an observational study of 35 interacting galaxy pairs,  that show gas concentration in the inner regions, with a subsequent enhancement of the central star formation rate]{smith07}. \\
In the previous section, we have shown indeed that the star formation enhancement in the merging phase is a strongly inhomogeneous process, taking place mostly in the central kiloparsec. Here we want to describe in more details the gas dynamics during the different phases of the encounter.
To do this, we studied the occurence and evolution of asymmetries in the disk stellar distribution and the response of the gas component to their emergence.
To this aim, firstly, we Fourier-analyzed the disk surface density distribution, focusing in particular our attention to the amplitudes of the Fourier components m=1 and m=2. Then we studied the evolution of the gas lagrangian radii containing $10\%$, $25\%$, $50\%$ and $75\%$ of the total gas mass, in order to have a picture of gas inflows into the central regions.
Evidently, it is not possible to describe here the results of all the different encounters simulated, so we will describe in detail only two of the cases analyzed.

To Fourier-analyze the disk surface density distribution, we adopted  the following procedure:
\begin{itemize}
\item firstly, the angular momentum of the disk galaxy has been evaluated and the galactic plane has been rotated consequently, in order to have the galactic spin parallel to the z-axis of the reference frame;
\item then the x-y plane has been divided into an annular grid (exponentially spaced in radius and linearly in azimuth) between the center $C$ of the disk galaxy and $10 kpc$.
\item the surface mass density distribution on the x-y plane has been evaluated, taking into account only the galactic old stellar component that lies between  $\pm 500$ pc from the galactic disk.
\item once  the density $\Sigma(r,\theta)$, being $r$ the distance from the galaxy center and $\theta$ the azimuth, is computed, for each annular ring, we have fitted $\Sigma(r,\theta)$ with the following function
   \begin{eqnarray}
     \frac{f(\theta)}{A_0}&=&1+[A_1 sin(\theta+\phi_1)+A_2sin(2\theta+\phi_2)+ \nonumber \\
   & &    ....+A_8sin(8\theta+\phi_8)]
   \end{eqnarray}
obtaining a value of $A_0$, $A_1$,...,$A_8$ and $\phi_1$,...,$\phi_8$ for each annular ring.
\item finally, for each configuration, we have evaluated the averaged value $<A_1>$ and $<A_2>$ of $A_1$ and $A_2$ respectively, between $1$ and $10$ kpc and we adopted these values to quantify the ``strength'' of the m=1 and m=2 asymmetries.
\end{itemize}

The results of this analysis are shown in Fig.\ref{inflow}, which refers to two encounters (a direct and a retrograde one) between an elliptical and a gSb galaxy. In the first column, the relative distance between the two interacting system is shown, as a function of time, while the second and third column show, respectively, the evolution of  m=1 and m=2 asymmetries. It is evident from this plot that: 

\begin{itemize}
\item usually tidal encounters amplify both m=1 and m=2 asymmetries;
\item at the first pericenter passage or during flybys, usually galaxies in direct orbits develop $A_2$ asymmetries more pronounced than those arising in retrograde encounters;
\item for mergers, a second amplification in the $A_1$ and $A_2$ asymmetries, relative to the isolated case, occurs when the two galaxies are in the final stage of coalescence.
\item in the phase between first pericenter passage and merger, the $A_1$ and $A_2$ values can be either stronger or lower than those of the same galaxies evolved in isolation
\end{itemize}
\begin{figure}
  \centering
  \includegraphics[width=8cm,angle=270]{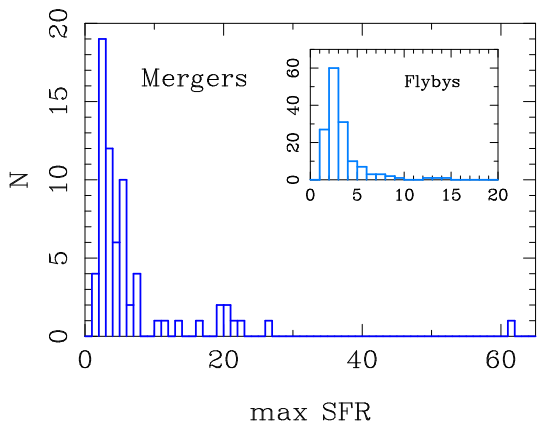}
  \caption{Histograms of the maximum SFR (relative to the isolated case), for mergers. Flybys are shown for comparison in the small window inserted in the Figure.
}
  \label{pisto}%
\end{figure}
\begin{figure}
  \centering
  \includegraphics[width=8cm,angle=270]{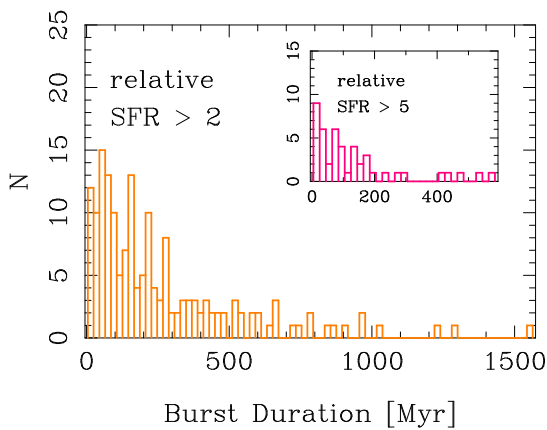}
  \caption{Histograms of duration of enhanced SFR, for the whole sample of interacting galaxies (mergers and flybys). Two thresholds are shown: relative SFR $>$ 2  and relative SFR $>$ 5 (the latter is shown in the small window inserted in the Figure).
}
  \label{pistotime}%
\end{figure}

The emergence of strong asymmetries in the disk obviously influences the gas dynamics, as shown in the last column of Fig.\ref{inflow}, where the lagrangian radii for  the gas component are shown\footnote{In fact here we refer to the gas+new stars component.}. Indeed, after the first pericenter passage, as the disk galaxy developes transient m=1 and m=2 asymmetries, a first gas inflow, involving about $25\%$ of the gas mass takes place, accompanied by an increase in the star formation efficiency. This inflow last about 200 Myr and it is followed by a re-expansion of the inner gas lagrangian radii, probably due to an enhancement in the gas kinetic energy from SNe explosions. But the most dramatic and very rapid gas compression starts only when the two interacting galaxies are in the final phase of the merger: in this case the rapidly changing galactic potential causes the inflow of a high gas mass fraction (between $50\%$ and $80\%$ of the gas mass being involved) into the inner kiloparsec region, where a strong starburst takes place.

%----------------
\section{Discussion: 
Are galaxy interactions always starburst triggers?}\label{discussion}

In the previous Section, we have investigated in details the evolution of the star formation rate during flybys and mergers, showing the great variety of SFR evolutions occurring during galaxy interactions, and presenting a deep study of the SFR dependence on several parameters. 
 We have already pointed out (Section \ref{where}) that interactions are not always a sufficient condition to convert high gas mass quantities into new stars. Here we want to deepen the discussion on this point, beacause of its potential impact also for observational studies.
Evidently, we keep in mind that the statistical analysis presented in this paper contains inevitable limitations, that we will try to remove in following works. Even exploring a whole Hubble sequence (from early-tape to late-type giant galaxies), all the simulations have been performed using an unique dark matter model, for example. Having shown in Section \ref{tidal} that tidal effects are crucial in determining the SFR in the merging phase, it could be interesting to check, in subsequent works, how the overall analysis depends on the dark matter model adopted (varying the density profiles, the limiting radii, the central concentrations, etc,...). \\

\begin{figure}
  \centering
  \includegraphics[width=8cm,angle=270]{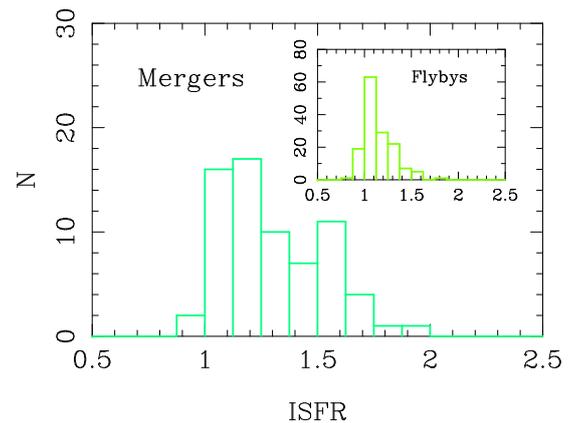}
  \caption{Histogram of the relative ISFR, for mergers.  Flybys are shown for comparison in the small window inserted in the Figure.
}
  \label{pistoisfr}%
\end{figure}
The present work is also limited to coplanar encounters, with galaxies having null relative disk inclinations, and spins parallel to the orbital angular momentum.\\
Anyway, the large sample (216) of interacting galaxies here presented is yet sufficient to draw some preliminary conclusions about the interaction-starburst connection.\\
Fig.\ref{pisto} shows the histogram of the maximum SFR for mergers (upper panel) and flybys (lower panel). As in the previous Section, this value is relative to the SFR of the galaxy evolved in isolation. The plot clearly puts in evidence that \emph{mergers are not always starburst triggers}: indeed the number of mergers that produces star formation rates 10 times higher than those of isolated galaxy are about the 17$\%$ of the total merger sample, while half of the sample shows only a moderate enhancement in the SFR (less than 4 times the isolated case).\\
High SFR  are not only less frequent, but are also characterized by shortest duration times. This is shown in Fig.\ref{pistotime}, where the histogram of the burst duration is plotted, for two different SFR enhancement levels. About 36$\%$ of the whole interacting galaxies sample\footnote{i.e. including flybys and mergers together.} sustains SFRs two times higher than those of the isolated counterparts for a time greater than 500 Myr (upper panel in Fig.\ref{pistotime}). But this  duration significatively diminishes, for galaxies whose relative SFR is higher than 5. In this case, the maximum starburst duration is less than 600 Myr, and only $13\%$ of the whole galaxy sample sustains a SFR enchancement greater than 5, for more than 100 Myr (lower panel in Fig.\ref{pistotime}). Finally galaxies that show SFRs 20 times higher than those of the isolated counterparts are able to substain it for 150 Myr, in the most favourable cases. \\
Not only the starbust frequence is low, but also the integrated star formation rate, compared to the isolated case, seems to support the idea that \emph{interactions are not always sufficient to convert high gas mass  quantities into new stars}, as the histogram in Fig.\ref{pistoisfr} shows. Flybys and mergers can produce 2 times more stars than isolated galaxies, but about $50\%$ of all the mergers, and $76\%$ of flybys, has an ISFR (see Section \ref{isfr} for its definition) which is only 1.25 higher than that of isolated galaxies.

%----------------
\section{Conclusions}\label{concl}
We have investigated the enhancement of the star formation rate in galaxy interactions, by numerical simulations, comparing the star formation properties of more than two hundred pairs of interacting and merging galaxies, with those of isolated galaxies.\\
The present work partially confirms previous numerical investigations \citep{mih94, mih94b, mih96, spri00, cox06}, in the sense that galaxy major interactions and mergers can trigger strong nuclear starburst, but it clearly puts in evidence that this is not always the case, i.e. %\begin{itemize}
%\item 
\emph{mergers are not always starburst triggers}
%\item 
and \emph{galaxy interactions are not a sufficient condition to convert high gas mass quantities into new stars}. 
%\end{itemize}
This mainly because strong tidal interactions at the first pericenter passage can remove a large amount of gas from the galaxy disks. This gas material, ejected into the tidal tails, is only partially re-acquired by the galaxies in the last phase of the merging event. \\

We have shown that the star formation rate in the merging phase:
\begin{enumerate}

 \item does not depend mainly on the total amount of gas mass available just before the final coalescence phase, being this total amount of gas content the sum of that present in the tails and in the main body of the system.\\
In turn, it is strongly anti-correlated with: 
 \item the distance of the two galaxy centers at the first pericenter passage, in the sense that, on average, galaxies that suffer too close passages produce also the lowest bursts of star formation;
 \item the  amplitude of the tidal forces at pericenter, i.e. pairs that suffer less intense tidal actions at the first passage are able to preserve a great mass fraction in the disk, that constitutes the fuel for the nuclear starburst in the merging phase. Furthermore,
\item the enhancement in the star formation in the merging phase  depends on the galaxy spin, in the sense that, on average, galaxies in retrograde orbits are stronger starburst triggers than those involved in direct encounters. 
\end{enumerate}

We have also analyzed the evolution in the  $(\Sigma_{gas}, \Sigma_{SFR})$ plane of interacting, mergers and post-merger galaxies, finding that, globally, the Kennicutt-Schmidt law is retrieved statistically for all the different stages of interaction.\\

Finally, we have proposed a formulation for the SFE at the pericenter passage, for flybys, and in the coalescence phase for mergers. The general laws we derived depend only on the main parameters of the encounters, as the orbital spin, the pericenter separation and the galaxy dimensions.\\

Obviously, it is still necessary to exploit a larger range of parameters, for example to get an insight into the dependence of the star formation efficiency on the orientation of the galaxy disks with respect to the orbital plane, or on the masses ratio of the galaxies in the pair. 
It would be interesting also to perform a subset of these simulations varying the star formation rule, i.e. from a density-dependent Schmidt type one, to a formulation that could take into account energy dissipation in shocks, as proposed by \citet{barn04}.
In any case, in our opinion, this work can contribute, on one hand, to clarify the physical mechanisms behind the large interval of star formation enhancements found in observed interacting pairs, an to furnish, on the other hand, simple star formation formulations for theoretical modelling.

\begin{acknowledgements}
P. D. M. thanks C. Jog for useful discussions and suggestions, and Y. Revaz for providing the python parallelized pNbody package (see  \emph{http://aramis.obspm.fr/$\sim$revaz/pNbody/index.html}), adopted to realize galaxy maps. \\
The authors are deeply grateful to F. Bournaud, for a careful reading of a first version of this paper, all his suggestions having undoubtfully improved the quality of the manuscript and the presentation of the results.  \\
We wish to thank an anonymous referee for his/her comments, that helped in improving the contents of this paper. \\

This research used computational resources of the Informatic Division of the Paris Observatory, and those available in the framework of the Horizon Project (see \emph{http://www.projet-horizon.fr/}).
\end{acknowledgements}

\bibliographystyle{aa} % style aa.bst
\end{document}